\documentclass[%
aps,
pra,
superscriptaddress,
tightenlines,
showpacs,showkeys,
a4paper,
12pt,
longbibliography,
notitlepage
]{revtex4-1}
\usepackage[english]{babel}
\usepackage{amssymb,amsmath,stmaryrd,array}

\usepackage{graphicx}

\makeatletter

\newcommand{\sca}[2]{\ensuremath{\bigl({#1}\cdot{#2}\bigr)}}
\newcommand{\avr}[1]{\ensuremath{\langle{#1}\rangle}}

\newcommand{\cnj}[1]{{#1}^{\ast}}
\newcommand{\hcnj}[1]{{#1}^{\dagger}}
\newcommand{\tcnj}[1]{{#1}^{T}}

\newcommand{\pdrs}[1]{\partial_{#1}}

\newcommand{\diag}{\mathop{\rm diag}\nolimits}

\newcommand{\mum}{$\mu$m}
\newcommand{\dega}{\ensuremath{^\circ}}
\newcommand{\degc}{$^\circ$C}

 \newcommand{\bs}[1]{\boldsymbol{#1}}
 \newcommand{\vc}[1]{\mathbf{#1}}
 \newcommand{\mvc}[1]{\mathbf{#1}}
 \newcommand{\uvc}[1]{\hat{\mathbf{#1}}}
 
 \newcommand{\ind}[1]{\mathrm{#1}}

\newcommand{\oprt}[1]{\ensuremath{\widehat{\mathcal{#1}}}}

\newcommand{\dd}{\mathrm{d}}

\newcommand{\eff}{\mathrm{eff}}

\newcommand{\inc}{\mathrm{inc}}
\newcommand{\refl}{\mathrm{refl}}
\newcommand{\transm}{\mathrm{trm}}
\newcommand{\vac}{\mathrm{vac}}
\newcommand{\med}{\mathrm{m}}

\makeatother

\begin{document}
\DeclareGraphicsExtensions{.eps,.png,.pdf}
\title{
Polarization gratings in the short-pitch approximation:
electro-optics of
deformed helix ferroelectric liquid crystals
}

\author{Alexei~D.~Kiselev}
\email[Email address: ]{kiselev@iop.kiev.ua}
\affiliation{%
 Institute of Physics of National Academy of Sciences of Ukraine,
 prospekt Nauki 46,
 03028 Ky\"{\i}v, Ukraine}
\affiliation{%
 Bogolyubov Institute for Theoretical Physics of National Academy of Sciences of Ukraine,
 Metrolohichna Street 14-b,
 Ky\"{\i}v, Ukraine}

\author{Eugene~P.~Pozhidaev}
\email[Email address: ]{epozhidaev@mail.ru}
 \affiliation{%
P.N. Lebedev Physics Institute of Russian Academy of Sciences,
Leninsky prospect 53, 117924 Moscow, Russia
 }

 \author{Vladimir~G.~Chigrinov}
 \email[Email address: ]{eechigr@ust.hk}
\affiliation{%
 Hong Kong University of Science and Technology,
 Clear Water Bay, Kowloon, Hong Kong
 }

 \author{Hoi-Sing~Kwok}
\email[Email address: ]{eekwok@ust.hk}
\affiliation{%
 Hong Kong University of Science and Technology,
 Clear Water Bay, Kowloon, Hong Kong
 }

\date{\today}

\begin{abstract}
Electro-optical properties
of deformed helix ferroelectric liquid crystal (DHFLC) cells
are studied by using a general theoretical approach to 
polarization gratings in which the transmission and reflection
matrices of diffraction orders are
explicitly related to the evolution operator
of equations for the Floqu\'et harmonics.
In the short-pitch approximation, a
DHFLC cell is shown to be optically equivalent to a uniformly
anisotropic biaxial layer where one of the optical axes is normal to the
bounding surfaces. 
For in-plane anisotropy, orientation of the optical axes
and birefringence are both
determined by the voltage applied across the cell
and represent the parameters that govern 
the transmittance of normally incident 
light passing through crossed polarizers.
We calculate the transmittance as a function of the
applied voltage and 
compare the computed curves with the experimental
data.
The theoretical and experimental results are found to be in good
agreement.
\end{abstract}


 \maketitle

\section{Introduction}
\label{sec:intro}

A \textit{polarization grating} (PG) can generally be described as
an optically anisotropic layer 
characterized by the anisotropy parameters that periodically vary
in space along a line in the plane of its input face.
Unlike conventional phase and amplitude diffraction gratings, PGs act by
locally modifying   the polarization state of light  waves passing
through them. 
Owing to the one-dimensional (1D) in-plane periodicity,
this introduces periodically modulated  changes of the
polarization characteristics giving rise to polarization-dependent
diffraction. 
In particular, the latter implies that a PG divides a monochromatic plane wave
into differently polarized diffracted waves.  

Over the past decade PGs have been attracted much
attention due to a unique combination of their optical properties: 
(a)~it is  possible to achieve 100\% diffraction into a single
order; 
(b)~diffraction efficiencies are highly sensitive to
the incident light polarization;
and (c)~the state of polarization of diffracted orders 
is determined solely by the parameters of a 
PG~\cite{Tod:1984,Gori:optl:1999,Tervo:optl:2000,Tervo:optcom:2001}.
There are numerous applications in a variety of fields,
including polarimeters, displays, polarizing beam splitters,
beam steering and polarization multiplexers
where PGs have been found to 
be useful
(for a recent review, see the article~\cite{Cincotti:ieee:2003}  and 
the monograph~\cite{Nikolova:bk:2009}).
 
There are different technologies to fabricate PGs.
For example,
computer-generated
subwavelength-period metal-stripe gratings with spatially periodic 
fringe orientation~\cite{Bomzon:optcom:2001}
and space-variant dielectric subwavelength 
gratings formed by discrete orientation of local subwavelength 
grooves~\cite{Bomzon:optl:1:2002,Biener:josa:2003}
are produced using advanced photolithographic and etching 
techniques.
Such gratings were employed to perform 
real-time polarization
measurements~\cite{Bomzon:optl:1:2002,Biener:josa:2003,Gorodetski:optl:2005}.
They are also used to demonstrate polarization Talbot
self-imaging~\cite{Bomzon:aplopt:2002}
and 
Pancharatnam-Berry phase optical elements~\cite{Bomzon:optl:2:2002,Niv:optl:2006}.
 
Polarization holography 
provides another well-known method to produce 
PGs~\cite{Nikolova:bk:2009}. 
It uses two differently polarized light beams
to record the spatially modulated 
polarization state of the resultant light field on suitable
media such as azobenzene containing polymer systems
and silver-halide materials.

The holographic technique has been extensively used to 
create polarization gratings in 
liquid crystal (LC) cells with 
photosensitive aligning substrates
such as
linear  photopolymerizable polymer layers~\cite{Eakin:apl:2004,Crawford:jap:2005}, 
azo-dye films~\cite{Presn:optex:2006},
azo-dye doped polyimide~\cite{Provenzano:apl:2006,Cipparrone:optex:2007}, 
and azobenzene side-chain polymer layers~\cite{Choi:josa:2009}.

In this method, 
irradiation of the substrate with a holographically  generated polarization 
interference pattern 
gives rise to spatially modulated light induced ordering
in the photoaligning layer.
This ordering  manifests itself in   
the effect of photoinduced optical anisotropy
and determines the anchoring properties
of the layer
such as its (polar and azimuthal) anchoring strengths
and the easy axis orientation
(see, e.g., Refs~\cite{Chigr:rewiev:2003,Kis:pre2:2005,Chigrin:bk:2008}
and references therein).

The anchoring parameters
of the photoaligning film 
 thus undergo periodic variations across 
the substrate face leading to the formation of  
orientational structures in the liquid crystal cells~\cite{Escuti:pre:2007} 
characterized by spatially periodic distributions
of the liquid crystal director, $\uvc{d}=(d_x,d_y,d_z)$,
which is a unit vector that defines a local
direction of the preferential orientation of LC molecules.
In liquid crystals,   
the elements of the dielectric tensor, $\bs{\varepsilon}$, can be expressed
in terms of the LC director~\cite{Gennes:bk:1993} 
\begin{align}
  \label{eq:diel_lc}
  \epsilon_{ij}=\epsilon_{\perp}(\delta_{ij}+
u_a\, d_{i}\, d_{j}), 
\quad
u_a=
(\epsilon_{\parallel}-\epsilon_{\perp})/\epsilon_{\perp},
\end{align} 
where $\delta_{ij}$ is the Kronecker symbol,
$u_a$ is the \textit{anisotropy parameter} and
$n_{\perp}\equiv n_o =\sqrt{\mu \epsilon_{\perp}}$
($n_{\parallel}\equiv n_e =\sqrt{\mu \epsilon_{\parallel}}$) 
is ordinary (extraordinary) refractive index
(the magnetic tensor of LC 
is assumed to be isotropic with the magnetic permittivity $\mu$). 
So, the periodic orientational LC configurations
define the so-called \textit{liquid crystal
polarization gratings} (LCPG). 
These gratings will be of our primary interest.
 
Common methods most generally employed to
derive theoretical results for PGs 
typically rely on on the well-known Jones matrix formalism 
and its 
modifications~\cite{Tod:1984,Gori:optl:1999,Tervo:optl:2000,Tervo:optcom:2001,
Cincotti:ieee:2003,Nikolova:bk:2009,Bomzon:optcom:2001,Bomzon:optl:1:2002,
Biener:josa:2003,Choi:josa:2009}.
These results are limited by their assumptions
to large gratings periods and normal incidence.  
In addition, using Jones calculus
implies neglecting multiple reflections.

\begin{figure*}[!tbh]
\centering
\resizebox{120mm}{!}{\includegraphics*{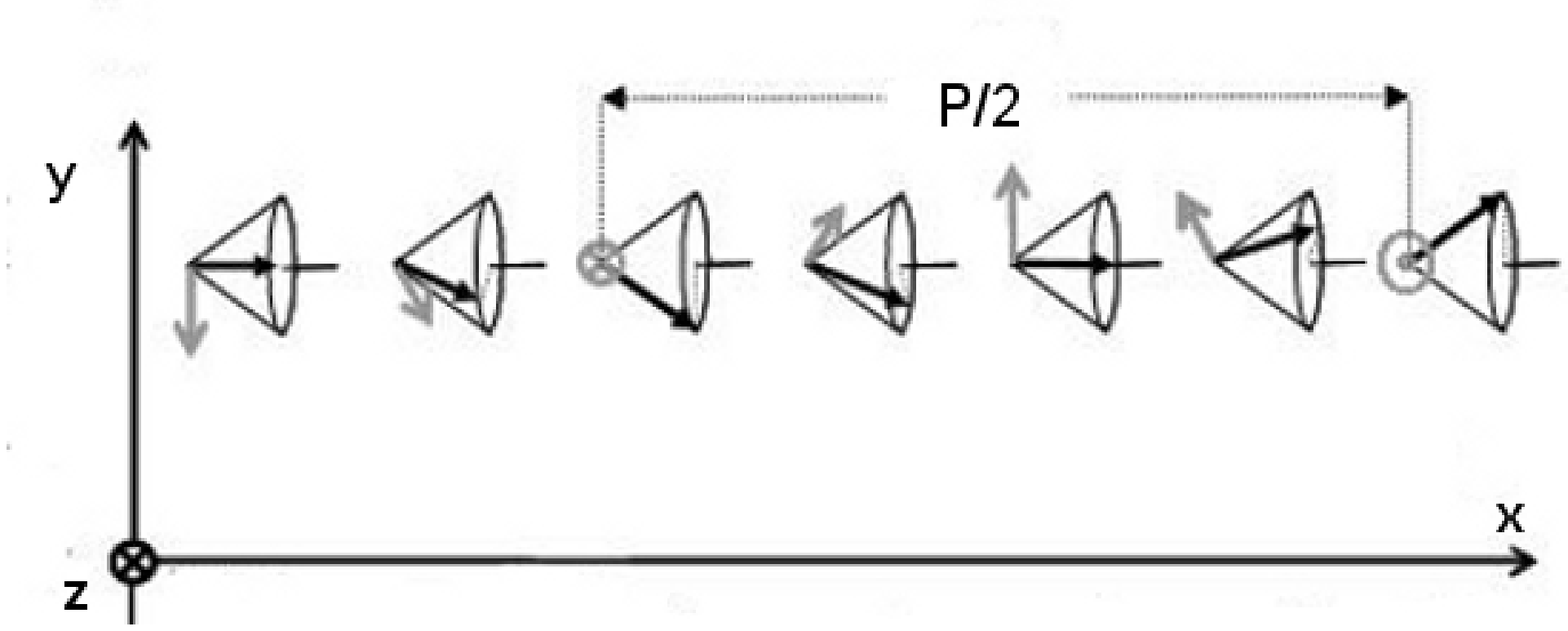}}
\caption{%
Helical structure of deformed helix ferroelectric liquid crystals.
The director rotates around the helix axis (the $x$ axis)
with cone angle $\theta_t$. 
}
\label{fig:dhf_struct}
\end{figure*}

In this work we present the theoretical approach to PGs that
can  be regarded as a generalized version of our method
developed in Refs.~\cite{Kis:jpcm:2007,Kiselev:pra:2008} for 
stratified anisotropic media
and goes beyond the limitations of Jones calculus. 
We apply the method for systematic treatment of
the technologically important case of the deformed helix ferroelectric
(DHF) liquid crystals~\cite{Beresnev:lc:1989,Chigr:1999} 
where the director rotates about a uniform
twist axis parallel to the substrates forming 
the ferroelectric LC (FLC) helical structure (see Fig.~\ref{fig:dhf_struct}).

In DHFLC cells, the FLC helix
is characterized by
a short submicron \textit{helix pitch}, $P< 1$~\mum, 
and a relatively large \textit{tilt angle}, 
$\theta_t>30\dega$. 
Note  that, in the case of surface stabilized FLC cells,
the helix pitch of a FLC mixture is typically greater than
the cell thickness, so that 
the bulk chiral helix turned out to be unwound (suppressed)
by the boundary conditions at the substrates~\cite{Clark:apl:1980}. 
By contrast to this, a DHFLC helix pitch is 5-10 times smaller than
the thickness. This allows the helix to be retained within
the cell boundaries. 

Electro-optical response of DHFLC cells
exhibits a  number of peculiarities that make them
useful for LC devices such as
high speed spatial light modulators~\cite{Abdul:mclc:1991,Cohen:aplopt:1997} 
and colour-sequential liquid crystal displays~\cite{Hedge:lc:2008}.
So, in this study, our goal is to examine electro-optical properties of
DHF liquid crystals based on the general theoretical approach  
describing polarization gratings. 

The layout of the paper is as follows.

In Sec.~\ref{sec:theory} we 
begin with Maxwell's equations
for the lateral components of the electric and magnetic fields
and derive a set of equations for the Floqu\'et harmonics
representing diffracted waves.
The relations linking
the transmission and reflection matrices of 
diffraction orders and
the evolution operator of the system for the harmonics
are deduced in Sec.~\ref{subsec:comp-method}.

Electro-optical properties of DHFLC cells
with the subwavelength helix pitch 
are studied in Sec.~\ref{sec:DHF}.
Experimental details 
are given in  Sec.~\ref{subsec:experiment}
where we describe the samples
and the setup employed to perform measurements. 
In Sec.~\ref{subsec:theory},
the general theory of Sec.~\ref{sec:theory}
is used 
to examine how
the optical anisotropy parameters and the transmission
coefficients of the DHFLC cells depend on the applied
electric field.
In particular, it is found that
in the short-pitch approximation
the DHFLC polarization gratings
can be represented by
uniformly anisotropic biaxial layers.
For the electric field dependence of 
the light transmittance through the cell placed between
crossed polarizers,
the results of electro-optic measurements
are compared with
the theoretically computed curves
in Sec.~\ref{subsec:expt-res-model}.

Finally, in Sec.~\ref{sec:discussion},
we present the results and make some concluding remarks.
Technical details on derivation of Maxwell's equations for
the lateral (in-plane) components of electromagnetic field 
are relegated to Appendix~\ref{sec:math}.
Theory of stratified media is recapitulated
in Appendix~\ref{sec:strat-anis-media}.

\begin{figure*}[!tbh]
\centering
\resizebox{120mm}{!}{\includegraphics*{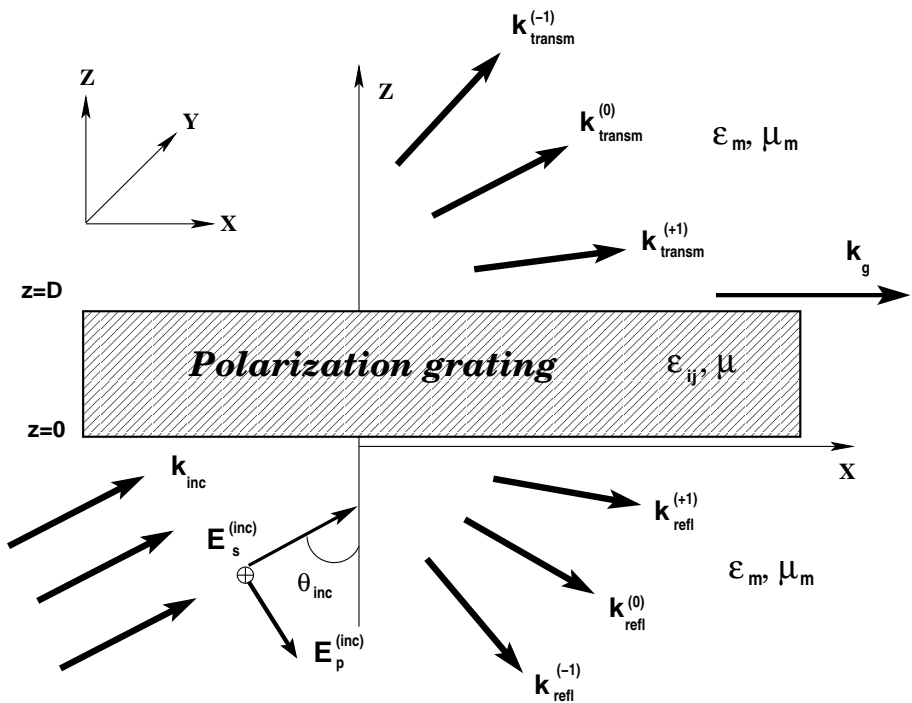}}
\caption{%
Polarization grating in the slab geometry. 
}
\label{fig:geom}
\end{figure*}

\section{Theory of polarization gratings}
\label{sec:theory}

In this section we
shall generalize the theoretical approach developed
in Refs.~\cite{Kis:jpcm:2007,Kiselev:pra:2008} 
so  as to treat the light transmission problem for 
a polarization grating
in the slab geometry illustrated in Fig.~\ref{fig:geom}.
In this geometry, 
as is indicated in Fig.~\ref{fig:geom},
the $z$ axis 
is normal to the bounding surfaces of the layer: $z=0$ and $z=D$,
the grating with the \textit{grating pitch}, $\Lambda_g$, 
and the \textit{grating wave vector},
\begin{align}
  \label{eq:grat_w_vector}
  \vc{k}_g=k_g\uvc{x}, 
\quad
k_g=\frac{2\pi}{\Lambda_g},
\end{align}
where $k_g$ is the  \textit{grating wave number},
 is characterized by the condition of  in-plane periodicity  
for the elements of the dielectric tensor,
$\bs{\varepsilon}$:
\begin{align}
\label{eq:grat_period}
 \epsilon_{ij}(x+\Lambda_g)= \epsilon_{ij}(x).
\end{align}
Expression for the grating wave vector~\eqref{eq:grat_w_vector} 
defines the $x$-$z$ plane as the \textit{plane of grating}.

Throughout this paper
we deal with harmonic electromagnetic fields
characterized by 
the \textit{frequency}, $\omega$, 
(time-dependent factor is $\exp\{-i \omega t\}$) 
and 
the \textit{free-space wave number}, $k_{\vac}=\omega/c$.
So, the starting point of our theoretical considerations
is the Maxwell equations for a harmonic
electromagnetic wave written in the form:
\begin{subequations}
  \label{eq:maxwell}
\begin{align}
&
  \label{eq:maxwell-E}
   \bs{\nabla}\times\vc{E}=i \mu k_{\vac} \vc{H},
\\
&
\label{eq:maxwell-H}
  \bs{\nabla}\times\vc{H}=-i k_{\vac}  \vc{D},
\end{align}
\end{subequations}
where 
$\vc{D}=\bs{\varepsilon}\cdot\vc{E}$ is the electric displacement field.

We shall also assume (see Fig.~\ref{fig:geom}) that
the medium surrounding the layer is
optically isotropic with 
the dielectric constant $\epsilon_{\med}$,
the magnetic permittivity $\mu_{\med}$ and the refractive index
$n_{\med}=\sqrt{\mu_{\med} \epsilon_{\med}}$. So, 
for a plane wave travelings along the wave vector
\begin{align}
  \label{eq:k_pl-w}
  \vc{k}=k_{\med}\uvc{k}=k_z\uvc{z}+\vc{k}_p,
\quad
\vc{k}_p=k_p \left[
\cos(\phi)\,\uvc{x}+\sin(\phi)\,\uvc{y}
\right],
\quad
k_{\med}=n_{\med}k_{\vac},
\end{align}
the electromagnetic field $\{\vc{E}, \vc{H} \}$ is given by
\begin{subequations}
\label{eq:pln-wv}
\begin{align}
&
  \label{eq:EH_pl-w}
  \{\vc{E}, \vc{H} \}=
\{\vc{E}(\uvc{k}), \vc{H}(\uvc{k})\}\exp[i \sca{\vc{k}}{\vc{r}}],
\\
&
\label{eq:E_pl-w}
\vc{E}(\uvc{k})=E_p\,\vc{e}_x(\uvc{k})+E_s\,\vc{e}_y(\uvc{k}),
\\
&
\label{eq:H_pl-w}
\frac{\mu_{\med}}{n_{\med}}\,\vc{H}(\uvc{k})=\uvc{k}\times\vc{E}(\uvc{k})=
E_p\,\vc{e}_y(\uvc{k})-E_s\,\vc{e}_x(\uvc{k}),
\end{align}
\end{subequations}
where the unit vectors $\uvc{k}=(\sin\theta\cos\phi,\sin\theta\sin\phi,\cos\theta)$,
$\vc{e}_x(\uvc{k})=(\cos\theta\cos\phi,\cos\theta\sin\phi,-\sin\theta)$
and $\vc{e}_y(\uvc{k})=(-\sin\phi,\cos\phi,0)$ expressed
in terms of the polar ($\theta$) and azimuthal ($\phi$) angles
form an orthogonal basis. 

The incoming incident
wave $\{\vc{E}_{\inc}, \vc{H}_{\inc} \}$ 
is represented by a plane wave~\eqref{eq:pln-wv},
$\{\vc{E}, \vc{H}\}=\{\vc{E}_{\inc}, \vc{H}_{\inc} \}$,  
propagating along the wave vector 
\begin{align}
  \label{eq:k_inc}
  \vc{k}=\vc{k}_{\inc}=k_z^{(\inc)}\uvc{z}+\vc{k}_p,
\quad
k_z^{(\inc)}=\sqrt{k_{\med}^2-k_p^2}
\end{align}
in the half space $z\le 0$
bounded by the input face of the grating.
In this case,
the polar angle, $0\le \theta=\theta_{\inc}<\pi/2$, 
is the \textit{angle of incidence}, whereas the azimuthal angle,
$0\le \phi=\phi_{\inc}< 2 \pi$, is the angle between the grating plane
and the \textit{plane of incidence}. 

\subsection{Maxwell's equations for lateral components}
\label{subsec:maxwll-lateral}

Now we write down
the representation for the electric and magnetic fields, $\vc{E}$ and
$\vc{H}$,
\begin{align}
  \label{eq:decomp-E}
  \vc{E}=E_z \uvc{z} +\vc{E}_{P},\quad
\vc{H}= H_z \uvc{z} +\uvc{z}\times \vc{H}_{P},
\end{align}
where the  components directed along the normal to the bounding surface
(the $z$ axis) are separated from the tangential (lateral) ones. 
In this representation,
the vectors
$\vc{E}_{P}=E_x \uvc{x}+E_y \uvc{y}\equiv
\begin{pmatrix}
  E_x\\E_y
\end{pmatrix}
$
and
$\vc{H}_{P}=\vc{H}\times\uvc{z}\equiv
\begin{pmatrix}
  H_y\\-H_x
\end{pmatrix}
$
are parallel to the substrates
and give the lateral components of the electromagnetic field.
Similar decomposition for the differential operator that enter
the Maxwell equations~\eqref{eq:maxwell} is
given by
\begin{align}
  \label{eq:decomp-nabla}
  k_{\vac}^{-1} \bs{\nabla}=\uvc{z}\,\pdrs{\tau} +i \bs{\nabla}_{p},
\quad
\bs{\nabla}_{p}^{\perp}= \uvc{z}\times \bs{\nabla}_{p},
\end{align}
where $\tau=k_{\vac} z$;
$
\bs{\nabla}_{p}=-i\, k_{\vac}^{-1}\,
(\uvc{x}\,\pdrs{x}+\uvc{y}\,\pdrs{y})\equiv
(\nabla_x,\nabla_y)
$
and
$
\bs{\nabla}_{p}^{\perp}=
(\nabla_x^{\perp},\nabla_y^{\perp})=(-\nabla_y,\nabla_x).
$

We can now substitute Eqs.~\eqref{eq:decomp-E} and~\eqref{eq:decomp-nabla}
into the system~\eqref{eq:maxwell}
and follow the algebraic procedure 
described in Appendix~\ref{sec:math}
to  eliminate the $z$-components of the electric and magnetic fields.
As a result, the $z$-components, $E_z$ and $H_z$,
turned out to be expressed in terms of the lateral components
(see Eqs.~\eqref{eq:H_z} and~\eqref{eq:E_z}).
The resultant system of differential equations for the lateral components,
$\vc{E}_P$ and $\vc{H}_P$,
can be written in the following matrix form:
\begin{align}
  \label{eq:op-system}
  -i\pdrs{\tau}\vc{F}=\oprt{M}\cdot\vc{F}\equiv
    \begin{pmatrix}\oprt{M}_{11}&\oprt{M}_{12}\\ \oprt{M}_{21}&\oprt{M}_{22} \end{pmatrix}
    \begin{pmatrix}\vc{E}_{P}\\\vc{H}_{P} \end{pmatrix},
\quad
\tau=k_{\vac} z,
\end{align}
where the elements of the matrix differential operators $\oprt{M}_{ij}$ are given by
\begin{subequations}
  \label{eq:M_ij_op}
\begin{align}
&
  \label{eq:M_11_12_op}
  \oprt{M}_{\alpha\beta}^{(11)}=
-\nabla_{\alpha}\cdot[\epsilon_{zz}^{-1}\epsilon_{z\beta}],
\quad
  \oprt{M}_{\alpha\beta}^{(12)}=
\mu\delta_{\alpha\beta}
-\nabla_{\alpha}\cdot\epsilon_{zz}^{-1}\cdot\nabla_{\beta},
\\
&
  \label{eq:M_22_21_op}
  \oprt{M}_{\alpha\beta}^{(22)}=
-\epsilon_{\alpha z}\epsilon_{zz}^{-1}\nabla_{\beta},
\quad
  \oprt{M}_{\alpha\beta}^{(21)}=
\epsilon_{\alpha\beta}^{(P)}
-\mu^{-1} \nabla_{\alpha}^{\perp}\cdot\nabla_{\beta}^{\perp}
\end{align}
\end{subequations}
and the elements of the effective dielectric tensor~\eqref{eq:diel-p}
that enter the operator $\oprt{M}_{\alpha\beta}^{(21)}$
are 
\begin{align}
  \label{eq:eps-p-ij}
 \epsilon_{\alpha\beta}^{(P)}=
\epsilon_{\alpha\beta}-
\epsilon_{\alpha z}\epsilon_{zz}^{-1}\epsilon_{z \beta},
\quad
\alpha,\beta\in\{x,y\}. 
\end{align}

\subsection{Floqu\'et harmonics}
\label{subsec:floquet-harm}

From Eq.~\eqref{eq:grat_period},
the dielectric tensor is a periodic function of $x$,
so that it is represented by  its Fourier series expansion
\begin{align}
  \label{eq:diel-Four-grt}
  \bs{\varepsilon}=
  \sum_{n=-\infty}^{\infty}\bs{\varepsilon}_{n}\exp[i  n\sca{\vc{k}_g}{\vc{r}_p}]
=
  \sum_{n=-\infty}^{\infty}\bs{\varepsilon}_{n}\exp(i  n k_g  x),
\end{align}
where $\vc{r}_p\equiv (x,y,0)$.
Similar remark applies to the coefficients
that enter the differential operators~\eqref{eq:M_ij_op}.

The representation of Floqu\'et harmonics~\cite{Kuchment:bk:1993} 
for solutions of the system~\eqref{eq:op-system}
\begin{align}
&
  \label{eq:Fn-grt}
\vc{F}(\vc{r})=
    \vc{F}(\vc{r}_p,\tau)=\sum_{n=-\infty}^{\infty}\vc{F}_{n}(\tau)
\exp\{i 
(\vc{k}_{n}\cdot\vc{r}_p)
   \},
\quad
\vc{k}_{n}=\vc{k}_{p}+n \vc{k}_g=k_{\vac}\vc{q}_n,
\\
&
\label{eq:q_n}
\vc{q}_n=q_n (\cos\phi_n\,\uvc{x}+\sin\phi_n\,\uvc{y})=
(q_x^{(n)},q_y^{(n)},0),
\end{align}
where   $\vc{k}_p$ is the tangential component of 
the wave vector of the incident plane wave defined in 
Eq.~\eqref{eq:k_inc},
is another consequence of the periodicity
condition~\eqref{eq:grat_period}. 

On substituting the expansion over the Floqu\'et harmonics~\eqref{eq:Fn-grt}
into the system~\eqref{eq:op-system}, we derive
a set of matrix equations for the Floqu\'et harmonics  
\begin{align}
&
  \label{eq:sys-Fn-grt}
  -i\pdrs{\tau}\vc{F}_{n}(\tau)=
\sum_{m=-\infty}^{\infty}\mvc{M}_{nm}(\tau)\cdot\vc{F}_{m}(\tau),
\\
&
\label{eq:Mnm-grt}
\mvc{M}_{nm}=
    \begin{pmatrix}\mvc{M}_{nm}^{(11)}&\mvc{M}_{nm}^{(12)}\\ \mvc{M}_{nm}^{(21)}&\mvc{M}_{nm}^{(11)} \end{pmatrix}
=
\frac{1}{\Lambda_g}\int_{0}^{\Lambda_g}
\exp\{-i 
(\vc{k}_{n}\cdot\vc{r})
   \}
\bigl[
\oprt{M}\exp\{i 
(\vc{k}_{m}\cdot\vc{r})
   \} 
\bigr]
\dd x,
\end{align}
where  the $2\times 2$ block matrices
$\mvc{M}_{nm}^{(ij)}$ are given by
\begin{subequations}
\label{eq:Mij-grt}
   \begin{align}
&
\label{eq:Mii-grt}
\left[\mvc{M}_{nm}^{(11)}\right]_{\alpha\beta}=-
q_{\alpha}^{(n)}\beta_{z \beta}^{(n-m)},
\quad
\left[\mvc{M}_{nm}^{(22)}\right]_{\alpha\beta}=-\beta_{\alpha z}^{(n-m)}
q_{\beta}^{(m)},
\\
&
\label{eq:M12-grt}
\left[\mvc{M}_{nm}^{(12)}\right]_{\alpha\beta}=\mu \delta_{\alpha
  \beta}\delta_{n m}- 
q_{\alpha}^{(n)}\, \eta_{zz}^{(n-m)}\,q_{\beta}^{(m)},
\\
&
\label{eq:M21-grt}
\left[\mvc{M}_{nm}^{(21)}\right]_{\alpha\beta}=\epsilon_{\alpha\beta}^{(n-m)}-
\mu^{-1}\delta_{nm}\, p_{\alpha}^{(n)}p_{\beta}^{(m)},\quad
\vc{p}_n=\uvc{z}\times\vc{q}_n,     
\end{align}
\end{subequations}
and $\eta_{zz}^{(n)}$,
$\beta_{ij}^{(n)}$ and $\epsilon_{\alpha\beta}^{(n)}$
are the Fourier coefficients for
$\epsilon_{zz}^{-1}$,
$\epsilon_{zz}^{-1}\epsilon_{ij}$
and $\epsilon_{\alpha\beta}^{(P)}$, respectively.

General solution of the system~\eqref{eq:sys-Fn-grt}
\begin{align}
  \label{eq:evol-oprt-grt}
  \vc{F}_{n}(\tau)=
  \sum_{m=-\infty}^{\infty}\mvc{U}_{nm}(\tau,\tau_0)\cdot\vc{F}_{m}(\tau_0)
\end{align}
can be conveniently expressed in terms of
the \textit{evolution operator} defined as the matrix solution of 
the initial value problem
\begin{subequations}
  \label{eq:evol_problem_grt}
\begin{align}
  \label{eq:evol_eq_grt}
     -i\pdrs{\tau}\mvc{U}_{nm}(\tau,\tau_0)
&
=
\sum_{k=-\infty}^{\infty}
\mvc{M}_{nk}(\tau)\cdot\mvc{U}_{km}(\tau,\tau_0),
\\
  \label{eq:evol_ic_grt}
\mvc{U}_{nm}(\tau_0,\tau_0)
&
=\mvc{I}_4\,\delta_{nm},
\end{align}
\end{subequations}
where $\mvc{I}_n$ is the $n\times n$ identity matrix.

The Floqu\'et harmonics, $\vc{F}_n$, represent the electromagnetic field
of the diffracted waves with the integer $n\in \mathbb{Z}$ 
giving the \textit{diffraction order}. From the system~\eqref{eq:sys-Fn-grt},
it can be inferred that the effect of inter-harmonics coupling 
responsible for diffraction is solely caused by the periodic modulation
of the dielectric tensor~\eqref{eq:diel-Four-grt}. 

In the ambient medium with $\epsilon_{ij}=\epsilon_{\med}\delta_{ij}$
and $\mu=\mu_{\med}$, the harmonics are decoupled
and, for the $n$th diffraction order, 
represent plane waves propagating along  
the wave vectors with the tangential component~\eqref{eq:q_n}.
The Floqu\'et harmonics of such waves, $\vc{F}_{n}^{(\med)}$, 
is given by
\begin{align}
&
  \label{eq:F_n-med}
  \vc{F}_n^{(\med)}(\tau)=\mvc{V}_{\med}(\vc{q}_n)
  \begin{pmatrix}
    \exp\{i \mvc{Q}_{\med}(q_n)\, \tau\} & \mvc{0}\\
\mvc{0} &    \exp\{-i \mvc{Q}_{\med}(q_n)\, \tau\}   
  \end{pmatrix}
  \begin{pmatrix}
    \vc{E}_{+}^{(n)}\\
\vc{E}_{-}^{(n)}
  \end{pmatrix},
\\
&
\label{eq:Q_med}
\mvc{Q}_{\med}(q_n)=q_{\med}(q_n)\,\mvc{I}_2,
\quad
q_{\med}(q_n)=\sqrt{n_{\med}^2-q_n^2},
\end{align}
where 
$\mvc{V}_{\med}(\vc{q}_n)$
is the eigenvector matrix for the ambient medium
given by
\begin{align}
&
\label{eq:Vm-phi-q}
\mvc{V}_{\med}(\vc{q}_n)=
\mvc{T}_{\ind{rot}}(\phi_n)\mvc{V}_{\med}(q_n)=
\begin{pmatrix}
 \mvc{Rt}(\phi_n)&\mvc{0}\\
\mvc{0}& \mvc{Rt}(\phi_n) 
\end{pmatrix}
\begin{pmatrix}
\mvc{E}_{\med} & -\bs{\sigma}_3 \mvc{E}_{\med}\\
\mvc{H}_{\med} & \bs{\sigma}_3 \mvc{H}_{\med}\\
\end{pmatrix},
\\
&
  \label{eq:EH-med}
  \mvc{E}_{\med}=
  \begin{pmatrix}
    q_{\med}(q_n)/n_{\med}& 0\\
0 & 1
  \end{pmatrix},
\quad
 \mu_{\med}\,\mvc{H}_{\med}=
 \begin{pmatrix}
   n_{\med}& 0\\
0 & q_{\med}(q_n)
 \end{pmatrix},
\\
&
\label{eq:Rot_matrix}
\mvc{Rt}(\phi)=\begin{pmatrix}
  \cos\phi &-\sin\phi\\
\sin\phi & \cos\phi
\end{pmatrix},
\end{align}
and $\{\bs{\sigma}_1,\bs{\sigma}_2,\bs{\sigma}_3\}$ are the Pauli matrices
\begin{align}
  \label{eq:pauli}
      \bs{\sigma}_1=
      \begin{pmatrix}
        0&1\\1&0
      \end{pmatrix},
\:
      \bs{\sigma}_2=
      \begin{pmatrix}
        0&-i\\i&0
      \end{pmatrix},
\:
      \bs{\sigma}_3=
      \begin{pmatrix}
        1&0\\0&-1
      \end{pmatrix}.
\end{align}

From Eq.~\eqref{eq:F_n-med},
the vector  amplitudes $\vc{E}_{+}^{(n)}$ and
$\vc{E}_{-}^{(n)}$ correspond to the forward and backward eigenwaves
with $k_z^{(+)}=+k_{\vac} q_{\med}$ and $k_z^{(-)}=-k_{\vac}
q_{\med}$, respectively.
 In the half space $z\le 0$, these describe 
the \textit{incident and reflected  waves}
 \begin{align}
   &
   \label{eq:inc}
   \vc{E}_{+}^{(n)}\vert_{z\le 0}=
\vc{E}_{\inc}^{(n)}=\delta_{n\, 0}\,\vc{E}_{\inc}
\equiv \delta_{n\, 0}
\begin{pmatrix}
E_{p}^{(\inc)}\\
E_{s}^{(\inc)}
\end{pmatrix},
\\
&
   \label{eq:refl}
   \vc{E}_{-}^{(n)}\vert_{z\le 0}=
\vc{E}_{\refl}^{(n)}\equiv
  \begin{pmatrix}
E_{p,\,n}^{(\refl)}\\
E_{s,\,n}^{(\refl)}
\end{pmatrix}.
 \end{align}
Clearly, equation~\eqref{eq:inc} implies that
the incident wave is the forward eigenwave of the zeroth order. 

In the half space $z\ge D$ after the exit face of the grating,
the only  wave of the $n$th order is 
the \textit{transmitted plane wave}
\begin{align}
   \label{eq:transm}
   \vc{E}_{+}^{(n)}\vert_{z\ge D}=
\vc{E}_{\transm}^{(n)}\equiv
  \begin{pmatrix}
E_{p,\,n}^{(\transm)}\\
E_{s,\,n}^{(\transm)}
\end{pmatrix},
\quad
   \vc{E}_{-}^{(n)}\vert_{z\ge D}=
\vc{0}.
 \end{align}
Note, that, at sufficiently large diffraction order with
$q_n>n_{\med}$ and $q_{\med}=i |q_{\med}|$ 
(see Eq.~\eqref{eq:Q_med}), 
the reflected and transmitted waves become
evanescent. 
In this case, the $z$-components of the wave vectors, 
$\vc{k}_{\refl}^{(n)}=k_{\vac}(-q_{\med}\,\uvc{z}+\vc{q}_n)$
and 
$\vc{k}_{\transm}^{(n)}=k_{\vac}(q_{\med}\,\uvc{z}+\vc{q}_n)$,
are imaginary.

\subsection{Computational procedure}
\label{subsec:comp-method}

We can now define
the \textit{transmission and reflection matrices}
through the linear input-output relations
\begin{align}
  \label{eq:transm-rel}
\vc{E}_{\transm}^{(n)}
=\mvc{T}_{n}
\cdot
\vc{E}_{\inc},
\quad
\vc{E}_{\refl}^{(n)}
=
\mvc{R}_{n}
\cdot
\vc{E}_{\inc}
\end{align}
linking the $n$the order for the transmitted and reflected waves 
to the incident wave
and, following the line of reasoning 
presented in Refs.~\cite{Kis:jpcm:2007,Kiselev:pra:2008},
relate these matrices and the evolution operator
given by Eq.~\eqref{eq:evol_problem_grt}.
To this end, we 
use  the boundary conditions requiring
the tangential components of the electric and magnetic
fields to be continuous at the boundary surfaces:
$\vc{F}_{n}(0)=\vc{F}_{n}^{(\med)}(0-0)$ and
$\vc{F}_{n}(h)=\vc{F}_{n}^{(\med)}(h+0)$,
and apply the relation~\eqref{eq:evol_problem_grt}
to  the anisotropic layer of the thickness $D$
to yield the following result
\begin{align}
  \label{eq:continuity}
  \vc{F}_{n}^{(\med)}(h+0)=\sum_{k}\mvc{U}_{nk}(h,0)\cdot\vc{F}_{k}^{(\med)}(0-0),
\quad  
h=k_{\vac} D.
\end{align}
By using Eq.~\eqref{eq:F_n-med},
equation~\eqref{eq:continuity} can be conveniently 
recast into the form
\begin{align}
&
\label{eq:linking}
  \begin{pmatrix}
\vc{E}_{\inc}^{(n)}\\ \vc{E}_{\refl}^{(n)}
\end{pmatrix}
=\sum_{k}
\mvc{W}_{nk}
\begin{pmatrix}
\vc{E}_{\transm}^{(k)}\\ \vc{0}
\end{pmatrix},
\end{align}
where $\mvc{W}_{nk}$  is given by
\begin{align}
&
\label{eq:W_nk}
   \mvc{W}_{nk}=
[\mvc{V}_{\med}(\vc{q}_n)]^{-1}\cdot[\mvc{U}^{-1}(h,0)]_{nk}\cdot\mvc{V}_{\med}(\vc{q}_k)=
\begin{pmatrix}
\mvc{W}_{nk}^{(11)} & \mvc{W}_{nk}^{(12)}\\
\mvc{W}_{nk}^{(21)} & \mvc{W}_{nk}^{(22)}
\end{pmatrix}. 
\end{align}
and defines the \textit{linking matrix}.

At $n\ne 0$, the relation~\eqref{eq:linking}
gives the system for the transmitted wave orders
\begin{align}
  \label{eq:sys-for-trm}
  -\mvc{W}_{n  0}^{(11)}\cdot\vc{E}_{\transm}^{(0)}= 
\sum_{k\ne 0} \mvc{W}_{n k}^{(11)}\cdot\vc{E}_{\transm}^{(k)},
\quad n\ne 0
\end{align}
By solving this system we obtain 
the diffracted waves
\begin{align}
  \label{eq:tilde-T_n}
  \vc{E}_{\transm}^{(n)}=\tilde{\mvc{T}}_{n}\cdot\vc{E}_{\transm}^{(0)}
\end{align}
linearly related  to the zeroth order (non-diffracted wave)
through the efficiency matrices $\tilde{\mvc{T}}_{n}$. 
Obviously, Eq.~\eqref{eq:tilde-T_n} implies that, by definition, 
$\tilde{\mvc{T}}_{0}=\mvc{I}_2$.

From Eq.~\eqref{eq:linking} at $n=0$, 
we obtain the relation 
\begin{align}
  \label{eq:sys-for-inc}
 \vc{E}_{\inc}= \sum_{k}\mvc{W}_{0 k}^{(11)}\cdot\vc{E}_{\transm}^{(k)}, 
\end{align}
linking the vector amplitudes of 
the incident and transmitted waves.
Substitution of the matrices from Eq.~\eqref{eq:tilde-T_n}
into Eq.~\eqref{eq:sys-for-inc}
provides the transmission matrix~\eqref{eq:transm-rel}
in the following form: 
\begin{align}
  \label{eq:Tn-gen}
  \mvc{T}_n=
\tilde{\mvc{T}}_{n}\cdot
\left[
\sum_{k}\mvc{W}_{0 k}^{(11)}\cdot\tilde{\mvc{T}}_{k}
\right]^{-1}.
\end{align}
From the relation~\eqref{eq:linking} applied to the case of reflected
waves, we derive the reflection matrix
\begin{align}
  \label{eq:Rn-gen}
  \mvc{R}_n=\sum_{k}\mvc{W}_{n k}^{(21)}\cdot \mvc{T}_k
\end{align}
expressed in terms of the transmission matrices~\eqref{eq:Tn-gen}.

In the limiting case of uncoupled harmonics, where
$\mvc{W}_{nk}=\delta_{nk}\mvc{W}_{nn}$, 
it is not difficult
to recover the results for stratified media obtained 
in Refs.~\cite{Kis:jpcm:2007,Kiselev:pra:2008}.
It will suffice to note that,
at $\mvc{W}_{0n}^{(11)}=\mvc{0}$,
the system~\eqref{eq:sys-for-trm} 
gives the efficiency matrices $\tilde{\mvc{T}}_n=\delta_{n0}\,\mvc{I}_2$
that point to the absence of diffracted waves.

\begin{figure*}[!tbh]
\centering
\resizebox{120mm}{!}{\includegraphics*{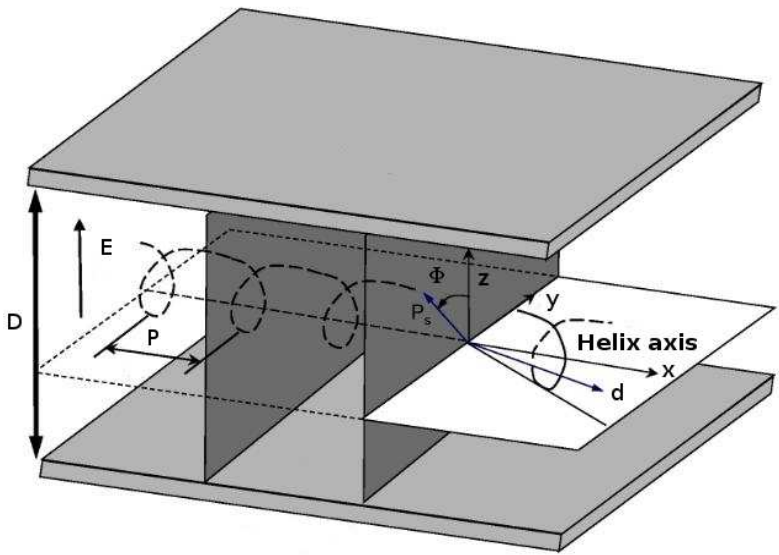}}
\caption{%
Geometry of a deformed helix FLC cell.  
}
\label{fig:dhf_cell}
\end{figure*}

\section{Deformed helix ferroelectric liquid crystal cells}
\label{sec:DHF}

In this section we present the experimental results
on the transmittance of light passing through crossed polarizers
measured as a function of the applied electric field (voltage)
in DHFLC cells. 
In order to interpret the experimental data,
the theoretical approach of Sec.~\ref{sec:theory}
is applied to chiral smectic helical
configurations with the subwavelength helix pitch. 

\subsection{Experiment}
\label{subsec:experiment}

\subsubsection{Sample preparation}
\label{subsubsec:sample}
 
The photo-alignment technique described in Ref.~\cite{Pozhidaev:jjap:2004} 
was used for producing the FLC cells. 
Geometry of the cells is schematically depicted in Fig~\ref{fig:dhf_cell}. 
We used
the cells with the size of $13\times 13$~mm$^2$, 
the thickness of the glass
substrate 1.1~mm, electrodes area $10\times 10$~mm$^2$, and the 
cell thickness (gap) 50~\mum\ and  130~\mum.  

Following the method of Ref.~\cite{Pozhidaev:jjap:2004}, 
ITO surfaces of FLC cells were
covered with a 10-20~nm photo-aligning substance - azobenzene sulfonic
dye SD-1 layers. 
The azo-dye solution was spin-coated onto ITO
electrode and dried at 155~\degc. 

The surface of the coated film was illuminated with linearly polarized
UV light using a super-high-pressure Hg lamp through an interference
filter at the wavelength 365~nm and a polarizing filter. 
The intensity of light irradiated normally on the
film surface during 30 minutes was 6~mW/cm$^2$.

In our experiments we used the FLC mixture FLC-576A 
(from
P. N. Lebedev Physical Institute of Russian Academy of Sciences)
 as a material for the DHFLC layer.
The mixture was injected into the cells in the isotropic phase 
by capillary action.
This mixture has the helix pitch $P\approx 300-330$~nm. 
Hence, in visible spectral range, the light
scattering  turns out to be
completely suppressed~\cite{Pozhidaev:mclc:2009}
except that the applied voltage is close to 
the critical voltage of the helix unwinding. 
So, such mixtures exhibit pronounced effects of
electrically controlled phase retardation and
birefringence $\Delta n_{\eff}(V)$ that can be clearly detected.

\begin{figure*}[!tbh]
\centering
\resizebox{120mm}{!}{\includegraphics*{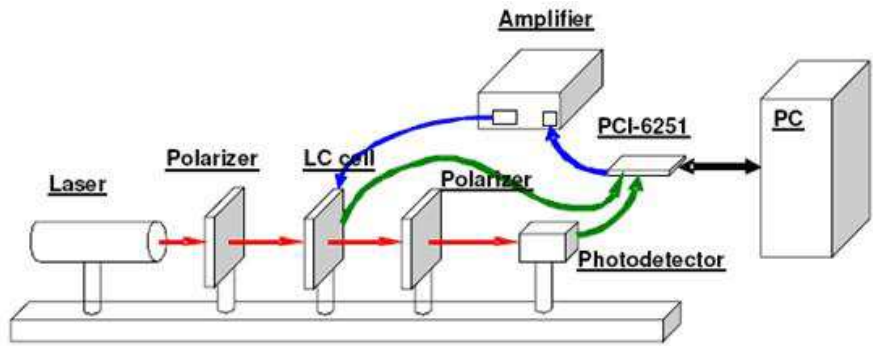}}
\caption{%
Experimental set-up for electro-optical measurements of 
DHFLC cells. 
}
\label{fig:setup}
\end{figure*}

\subsubsection{Experimental setup}
\label{subsubsec:setup}

The measurements of the light transmittance 
in relation to the applied voltage
were performed in the automatic regime. 
For this purpose, 
the measurement complex device was built 
whose principal scheme is shown in Fig.~\ref{fig:setup}. 
The basic element of this experimental set-up is computer
data acquisition (DAQ) board NIPCI 6251 from National
Instruments. This board has two analog outputs and 16 analog
inputs. The operating voltage is $\pm 10$~V, the maximal registration speed
is 1~$mu$s. The board has independent output and input buffer for 4000
point. In our experiments, the output signal 10~V was not sufficient and
the Wideband Power Amplifier KH model 7600 from Krohn-Hite Corporation
with amplification coefficients  5 and  25 times was used. 
It gives a possibility to have the output signal $\pm 250$~V. 
For the input signal this board has internal amplifier with coefficients
1, 2, 4, 8, 16, 32, 64, and 128 times. A photo-detector was connected
to input board plate for optical measurements.

The software for experimental set-up has built-in functions for analog
output-input. The program has three functional blocks that make 
operations with the set-up highly effective. The first block is a
programmable generator that realizes any form of signal with duration
of 2000 points. The duration of one point can be set from 1~$\mu$s to 
1~s. The second block is a measuring block, which save 4000 values of
the input voltage with step from 1~$\mu$s to 1~s. The operation of the
first and the second blocks is synchronized inside the DAQ board and
cannot be disturbed by computer interruptions. The third block is
used to accumulate the experimental data during the working period.

\subsection{Theory}
\label{subsec:theory}

In this section we apply our general theoretical approach
described in Sec.~\ref{sec:theory}
to the special case of DHFLC polarization gratings.
As is schematically shown in Figs.~\ref{fig:dhf_struct} and~\ref{fig:dhf_cell},
in deformed helix ferroelectric (DHF) liquid crystals, 
the director 
\begin{align}
&
\label{eq:dhf-director}
\uvc{d}=(d_x,d_y,d_z)=
\cos\theta_{t}\,\uvc{x}+
\sin\theta_{t}\,\cos\Phi\,\uvc{y}+
\sin\theta_{t}\,\sin\Phi\,\uvc{z}
\end{align}
lies on the smectic cone
with the \textit{helix tilt angle} $\theta_t$ 
and rotates in a helical fashion about a uniform in-plane
twist axis (the $x$ axis) forming the DHF helix.

In the presence of weak electric field, 
$\vc{E}=E\,\uvc{z}$, which is well below  its critical value
at the helix unwinding transition, $E\ll E_c$, 
the azimuthal angle around the cone, $\Phi$,
can be written in the form~\cite{Abdul:mclc:1991,Hedge:lc:2008}
\begin{align}
&
\label{eq:Phi}
 \Phi=\phi+\Delta\Phi(\phi)\approx \phi+\alpha_{E}\sin\phi,
\quad
\phi=\frac{2\pi}{P}\,x,  
\end{align}
where  the electric field induced
distortion $\Delta\Phi$ linearly depend
on $E$ through the \textit{electric field parameter}
$\alpha_{E}$ proportional to the ratio of the applied and
critical electric fields: $E/E_c$.

From Eq.~\eqref{eq:Phi}, it is clear that 
in the regime of weak electric field, 
the \textit{helix pitch} $P$ defines
both the grating period, $\Lambda_g=P$,
and the grating wave number, $k_g=2\pi/P$.
For the LC dielectric tensor~\eqref{eq:diel_lc}
with the DHF helical configuration~\eqref{eq:dhf-director}, 
the anisotropy parameters that enter
the matrices $\mvc{M}_{mn}$ 
whose block structure is described in Eq.~\eqref{eq:Mij-grt}
are given by
\begin{align}
&
  \label{eq:eta_zz}
\epsilon_{\perp}\eta_{zz}=\frac{1}{1+u_a d_z^2}=\frac{1}{1+v
  \sin^2\Phi},
\quad
v=u_a\sin^2\theta_t,
\\
 &
  \label{eq:beta_az}
\beta_{\alpha z}=\beta_{z \alpha}=
\frac{u_a d_z d_{\alpha}}{%
1+v\sin^2\Phi},
\\
&
\label{eq:epsilon_p}
\epsilon_{\perp}^{-1}\epsilon_{\alpha\beta}^{(P)}=
\delta_{\alpha\beta}
+\frac{u_a d_{\alpha} d_{\beta}}{%
1+v\sin^2\Phi}.
\end{align}

According to the computational procedure
developed in the previous section,
after inserting the above parameters into 
the matrices of 
the system~\eqref{eq:sys-Fn-grt}
for the Floqu\'et harmonics, we need to
 find the evolution operator by solving
the initial value problem~\eqref{eq:evol_problem_grt}.
Then computing the transmission and the reflection
matrices~\eqref{eq:transm-rel}
involve the following steps:
(a)~evaluation of the linking matrix~\eqref{eq:W_nk};
(b)~calculation of the efficiency matrices~\eqref{eq:tilde-T_n}
by solving the system of linear equation~\eqref{eq:sys-for-trm};
(c)~computing the matrices $\mvc{T}_n$  and $\mvc{R}_n$
from the formulas~\eqref{eq:Tn-gen}
and~\eqref{eq:Rn-gen}, respectively.

As a consequence of the anisotropy induced mode coupling,
there are an infinite number of coupled matrix equations
in the system~\eqref{eq:sys-Fn-grt}.
So, the evolution operator cannot be generally computed in the closed form 
without  resorting to the methods of the perturbation theory 
or numerical analysis.

In this paper, we shall restrict ourselves to the case where
the helix pitch is smaller than the wavelength, 
$P<\lambda$,
so as to interpret the experimental data measured
in DHFLC cells with the subwavelength pitch. 
In our experiments, 
the incident light travels in the air  
with  and normally impinges on the cell.
At $n_{\med}=1$ and $k_p=0$, the condition
\begin{align}
  \label{eq:subwvlength}
  q_g\equiv k_g/k_{\vac}=\lambda/P>1
\end{align}
implies that $q_n>n_{\med}$. 
In this case, from the above discussion 
after Eq.~\eqref{eq:transm},  
there are no diffracted waves propagating in the air
and nonzero orders of the transmitted and reflected
beams are evanescent.
In the zero-order approximation, 
where $\mvc{M}_{mn}=\delta_{m0}\delta_{n0} \mvc{M}_{00}$,
the evanescent waves are neglected and
the DHFLC cell appears to be
effectively described as a uniformly anisotropic layer
characterized by the matrix
\begin{align}
  \label{eq:avr-M}
  \mvc{M}_{00}\equiv \avr{\mvc{M}}=
  \begin{pmatrix}
    \avr{\mvc{M}}_{11} & \avr{\mvc{M}}_{12} \\
\avr{\mvc{M}}_{21} & \avr{\mvc{M}}_{22}
  \end{pmatrix},
\end{align}
where $\displaystyle \avr{\dots}=(2\pi)^{-1}\int_o^{2\pi}\dots\dd\phi$.
The elements of the $2\times 2$ block matrices $\avr{\mvc{M}}_{ij}$
are given by
\begin{align}
&
  \label{eq:M_avr-ii}
  \avr{\mvc{M}}_{\alpha\beta}^{(11)}
=-q_{\alpha}^{(p)}\avr{\beta_{z \beta}},
\quad
\avr{\mvc{M}}_{\alpha\beta}^{(22)}=-\avr{\beta_{\alpha z}}
q_{\beta}^{(p)},
\\
&
  \label{eq:M_avr-12}
\avr{\mvc{M}}_{\alpha\beta}^{(12)}
=\mu \delta_{\alpha\beta}- 
q_{\alpha}^{(p)}\, \avr{\eta_{zz}}\,q_{\beta}^{(p)},
\\
&
  \label{eq:M_avr-21}
\avr{\mvc{M}}_{\alpha\beta}^{(21)}
=
\epsilon_{\alpha\beta}^{(\eff)}
-
\mu^{-1}\, p_{\alpha}^{(p)}p_{\beta}^{(p)},\quad
\vc{p}_p=\uvc{z}\times\vc{q}_p,      
\end{align}
where $k_{\vac} \vc{q}_p=\vc{k}_p$ and
$\epsilon_{\alpha\beta}^{(\eff)}=\avr{\epsilon_{\alpha\beta}^{(P)}}$
is the averaged effective dielectric tensor.

\subsubsection{Effective dielectric tensor}
\label{subsubsec:effect-diel-tens}

Our next step is to evaluate the averages that enter
the matrix~\eqref{eq:avr-M}. 
We note that, 
from Eq.~\eqref{eq:beta_az}  and the result
\begin{align}
 \label{eq:averages-4}
   \avr{\sin\Phi\,[1+v\sin^2\Phi]^{-1}}=
   \avr{\sin\Phi\cos\Phi\,[1+v\sin^2\Phi]^{-1}}=0,
\end{align}
the averages $\avr{\beta_{\alpha z}}$ are zero.
When the averaged matrix~\eqref{eq:avr-M}
at $\avr{\beta_{\alpha z}}=0$
is compared with the matrix of a uniformly
anisotropic layer (see, e.g., Eq.~(25) in Ref.~\cite{Kiselev:pra:2008}),
where $\avr{\beta_{\alpha z}}= 
\epsilon_{\alpha z}/\epsilon_{zz}$,
$\avr{\eta_{zz}} =1/\epsilon_{zz}$
and $\epsilon_{\alpha\beta}^{(\eff)}=\epsilon_{\alpha\beta}^{(P)}$,
it immediately follows that the effective anisotropic medium
is biaxial. In this medium, the diagonal element
\begin{align}
  \label{eq:eps_eff_zz}
      \epsilon_{zz}^{(\eff)}=\avr{\eta_{zz}}^{-1}
\end{align}
gives the principal value of the effective dielectric
tensor for
the optic axis normal to the cell
(parallel to the $z$ axis).

For our purposes, it will suffice to
expand the averages up to second order terms 
in the field parameter
\begin{align}
  &
  \label{eq:averages-1}
    \avr{[1+v\sin^2\Phi]^{-1}}\approx [1+v]^{-1/2}(1+ v \gamma_v^2
    \alpha_E^2),
\\
&
  \label{eq:averages-2}
   \avr{\cos^2\Phi\,[1+v\sin^2\Phi]^{-1}}\approx \gamma_v (1+ [1+v]^{1/2} \gamma_v
    \alpha_E^2),
\\
&
  \label{eq:averages-3}
   \avr{\cos\Phi\,[1+v\sin^2\Phi]^{-1}}\approx -\gamma_v \alpha_E,
\quad
\gamma_v\equiv (\sqrt{1+v}-1)/v,
\end{align}
so that the in-plane components of the dielectric tensor
are given by
\begin{subequations}
  \label{eq:epsl-eff}
\begin{align}
  &
  \label{eq:epsl-avr-xy}
    \epsilon_{xy}^{(\eff)}=\gamma_{xy} \alpha_E,
\quad
\gamma_{xy}
=-\cos\theta_t\sin\theta_t\gamma_v (\epsilon_{\parallel}-\epsilon_{\perp}),
\\
&
  \label{eq:epsl-avr-xx}
    \epsilon_{xx}^{(\eff)}=\bar{\epsilon}+\Delta\epsilon=\epsilon_{xx}^{(0)}+\gamma_{xx}\alpha_E^2,
\\
&
  \label{eq:epsl-avr-yy}
   \epsilon_{yy}^{(\eff)}=\bar{\epsilon}-\Delta\epsilon=\epsilon_{yy}^{(0)}+\gamma_{yy}\alpha_E^2,
\end{align}
\end{subequations}
where
\begin{align}
&
  \label{eq:espl0-ii}
\epsilon_{xx}^{(0)}
=\epsilon_{\perp}(1+(u_a-v) [1+v]^{-1/2}),
\quad
    \epsilon_{yy}^{(0)}
=
\epsilon_{\perp}[1+v]^{-1/2},
\\
&
\label{eq:gamma-ii}
\gamma_{xx}
=\epsilon_{\perp} (u_a-v) v [1+v]^{-1/2}\gamma_v^2,
\quad
\gamma_{yy}
=\epsilon_{\perp} v [1+v]^{1/2}\gamma_v^2, 
\end{align}
The relation
\begin{align}
&
  \label{eq:phi-d}
  \tan(2\phi_\dd)=\gamma_{xy}\alpha_E/\Delta\epsilon  
\end{align}
defines the azimuthal angle, $\phi_d$,
giving orientation of the in-plane optical axes.

Similar to the LC dielectric tensor~\eqref{eq:diel_lc},
the tensor~\eqref{eq:epsl-eff}
can be expressed in terms of the ``director''
$\uvc{d}_{\eff}=(d_{x}^{(\eff)},d_{y}^{(\eff)},0)=(\cos\phi_{\dd},\sin\phi_{\dd},0)$
and its eigenvalues
\begin{align}
&
\label{eq:epsl-pm}
\epsilon_{\pm}=n_{\pm}^{\,2}/\mu=
\bar{\epsilon}\pm\Delta\epsilon\sqrt{1+\tan^2(2\phi_\dd)}
\end{align}
as follows
\begin{align}
  \label{eq:epsl-eff-diag}
     \epsilon_{\alpha\beta}^{(\eff)}=
\epsilon_{-} \delta_{\alpha\beta}+
(\epsilon_{-}-\epsilon_{+}) d_{\alpha}^{(\eff)}d_{\beta}^{(\eff)}.
\end{align}

\subsubsection{Light transmittance}
\label{subsubsec:light-transm}

For the case of normal incidence,
the transmission and reflection matrices can be easily obtained
from the results of Refs.~\cite{Kis:pre:2009,Kiselev:jetp:2010}
in the limit of the wave vectors with 
vanishing tangential component 
(see also Eq~\eqref{eq:T_norm_pln} derived in Appendix~\ref{subsec:planar}). 
We shall need to write down the resultant expression 
for the transmission matrix
\begin{align}
&
  \label{eq:T-norm}
 \mvc{T}(\phi_{\dd})
\equiv
\begin{pmatrix}
  t_{xx}&t_{xy}\\
t_{yx}& t_{yy}
\end{pmatrix}
=
\frac{t_{+}+t_{-}}{2}\,
\mvc{I}_2+\frac{t_{+}-t_{-}}{2}\,\
\mvc{Rt}(2 \phi_{\dd})\cdot
\bs{\sigma}_3,
\\
&
 \label{eq:t-pm}
  t_{\pm}=\frac{1-\rho_{\pm}^2}{%
1-\rho_{\pm}^2\exp(2in_{\pm}h)
}\exp(i n_{\pm} h),
\quad
\rho_{\pm}=\frac{n_{\pm}/\mu-n_{\med}/\mu_{\med}}{%
n_{\pm}/\mu+n_{\med}/\mu_{\med}}.
\end{align}
When the incident wave is linearly polarized along the
$x$ axis (the helix axis),
the transmittance coefficient 
\begin{align}
&
  \label{eq:T-xy}
  T_{xy}=|t_{xy}|^2=\frac{|t_{+}-t_{-}|^2}{4}\,\sin^2(2\phi_\dd),
\quad
\sin^2(2\phi_\dd)=\frac{\alpha_{E}^2}{\alpha_{E}^2+(\Delta\epsilon/\gamma_{xy})^2}\,,
\end{align}
where $h=k_{\vac} D$ is the thickness parameter,
describes the intensity of the light passing through 
crossed polarizers.
Note that, under certain conditions such as $|\rho_{\pm}|\approx 1$, 
the transmittance~\eqref{eq:T-xy} can be approximated by simpler formula
\begin{align}
  \label{eq:T-xy-approx}
 T_{xy} \approx\sin^2(\delta/2)\,\sin^2(2\phi_\dd), 
\end{align}
where
$\delta=\Delta n_{\eff}\, h=(n_{+}-n_{-}) h$
is the difference in optical path of the ordinary and extraordinary
waves known as the \textit{phase retardation}.
 
\begin{figure*}[!tbh]
\centering
\resizebox{120mm}{!}{\includegraphics*{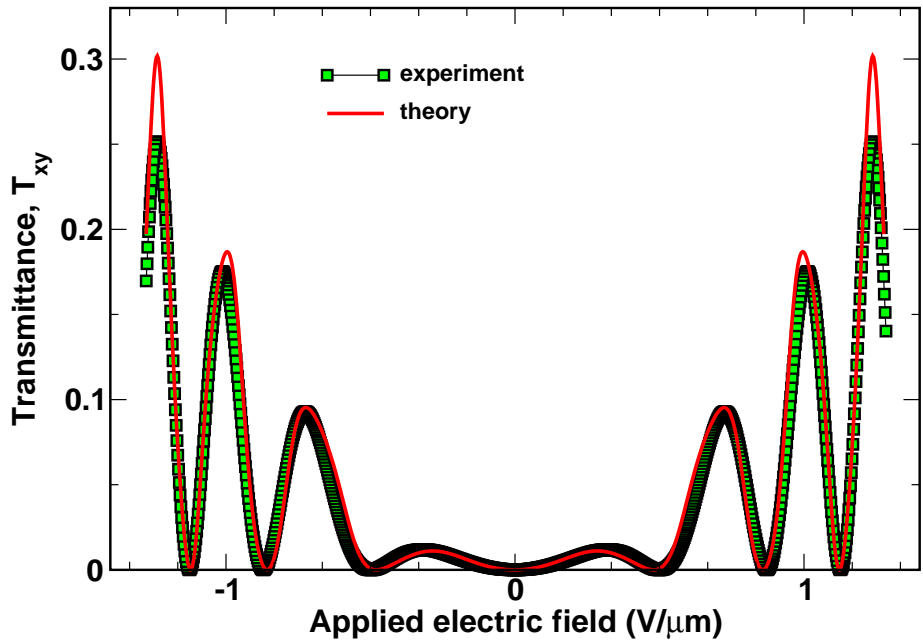}}
\caption{%
Transmittance of light passing through crossed polarizers, 
$T_{xy}$, as a function of the applied elecric
field for the DHFLC cell of thickness $D=130$~\mum\
filled with the FLC mixture FLC-576A.
Parameters of the mixture are: 
$n_o=1.5$ ($n_e=1.72$) is the ordinary (extraordinary)
refractive index and $\theta_t=32$~deg is the tilt angle.
The experimental points are marked by squares.
Solid line represents the theoretical curve
computed for the electric field parameter
$\alpha_{E}=\gamma_{E} E$ with
$\gamma_E\approx 0.62$~\mum/V.
}
\label{fig:data_130}
\end{figure*}

\subsection{Experimental results and modeling}
\label{subsec:expt-res-model}

Now we turn back to the electro-optical measurements
in the DHFLC cells described in Sec.~\ref{subsec:experiment}.
The transmittance versus electric field curve shown in Fig.~\ref{fig:data_130}
presents the results  measured at the wavelength of light
generated by the He-Ne laser with
 $\lambda=650$~nm in the cell in which
the thickness of the DHFLC layer, $D$,  was about 130~\mum.

The formula for the transmittance~\eqref{eq:T-xy}
can be combined with the expressions for the principal values of
the effective refractive indices~\eqref{eq:epsl-pm}
to evaluate dependence of $T_{xy}$ on the applied electric field, $E$.
The known parameters characterizing the FLC mixture FLC-576A 
that enter our formulas are the ordinary (extraordinary)
refractive index and the tilt angle
estimated at $n_o=1.5$ ($n_e=1.72$) and $\theta_t=32$~deg, 
respectively.
So, for the anisotropy parameters, $u_a$ and $v$, we have:
$u_a\approx 0.315$ and $v=u_a\sin^2\theta_t\approx 0.09$. 

From the discussion after Eq.~\eqref{eq:Phi},
the electric field parameter $\alpha_{E}$ is proportional
to the electric field, $\alpha_{E}=\gamma_{E} E$, 
and thus is determined by the coefficient of proportionality 
$\gamma_E$. This coefficient is the only fitting parameter
in our calculations.
Figure~\ref{fig:data_130} shows that
the theoretical curve computed at $\gamma_E\approx 0.62$~\mum/V
and the experimental data are in close agreement.

  \begin{figure*}[!tbh]
\centering
\resizebox{120mm}{!}{\includegraphics*{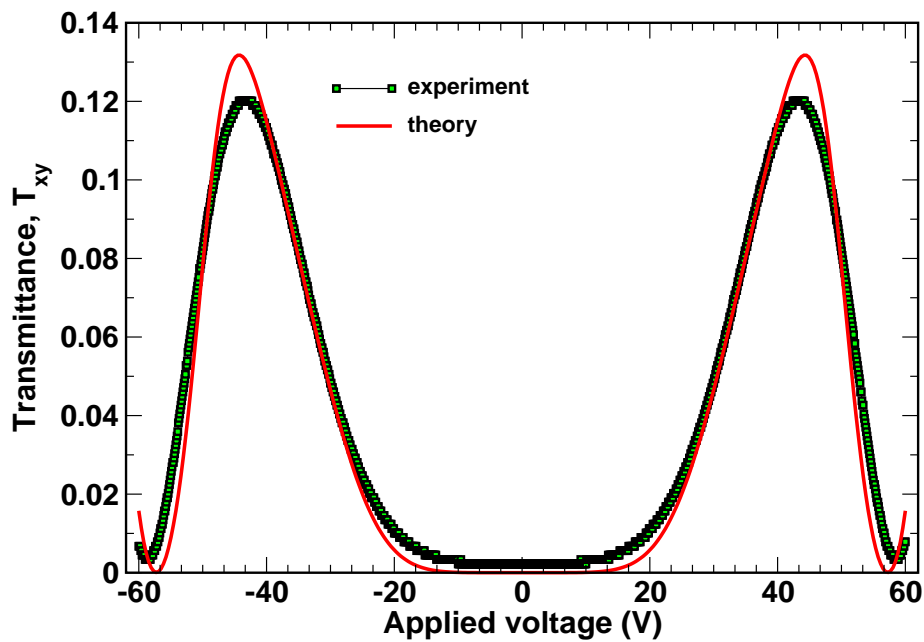}}
\caption{%
Light transmittance, 
$T_{xy}$, versus the voltage, $U$, applied across 
the DHFLC cell filled with the FLC mixture FLC-576A.
The cell thickness is estimated at about
$D=50\pm 2$~\mum.
The experimental points are marked by squares.
The theoretical curve (solid line) is
computed at the field parameter
$\alpha_{E}=\gamma_{V} U$ with
$\gamma_V\approx 0.0124$~V$^{-1}$.
}
\label{fig:data_50}
\end{figure*}

Similarly, in Fig.~\ref{fig:data_50},
 the transmittance $T_{xy}$ 
is plotted against the applied voltage, $U$,
for  the case where the cell thickness is 
about $50\pm 2$~\mum.
The theoretical results that give a good fit
to the data of measurements  
are evaluated
 at the field parameter $\alpha_{E}=\gamma_{V} U$ with
$\gamma_V\approx 0.0124$~V$^{-1}$.
So, in agreement with the above estimate, 
the cell thickness can be assessed at about 
$D=\gamma_E/\gamma_{V}\approx 50$~\mum. 

Referring to Figs~\ref{fig:data_130} and~\ref{fig:data_50}, 
the transmittance oscillates with the magnitude of 
the applied field. From Eqs.~\eqref{eq:t-pm} and~\eqref{eq:T-xy}, 
this is the dependence of the effective refractive indices,
$n_{+}$ and $n_{-}$, on the electric field parameter, $\alpha_{E}$,
that manifests itself in these oscillations.
More precisely, as is evident from the approximate expression
for the transmittance~\eqref{eq:T-xy-approx},
they are due to variations in the phase retardation, $\delta$,
arising from the electric field induced birefringence, 
$\Delta n_{\eff}=n_{+}-n_{-}$.
So, loci of extrema (minima and maxima) of the transmittance,
$T_{xy}$, 
can be used to obtain dependence of the birefringence on the applied voltage
(more details on this method can be found, e.g., in 
Ref.~\cite{Pozhidaev:lc:2010}). 

\begin{figure*}[!tbh]
\centering
\resizebox{120mm}{!}{\includegraphics*{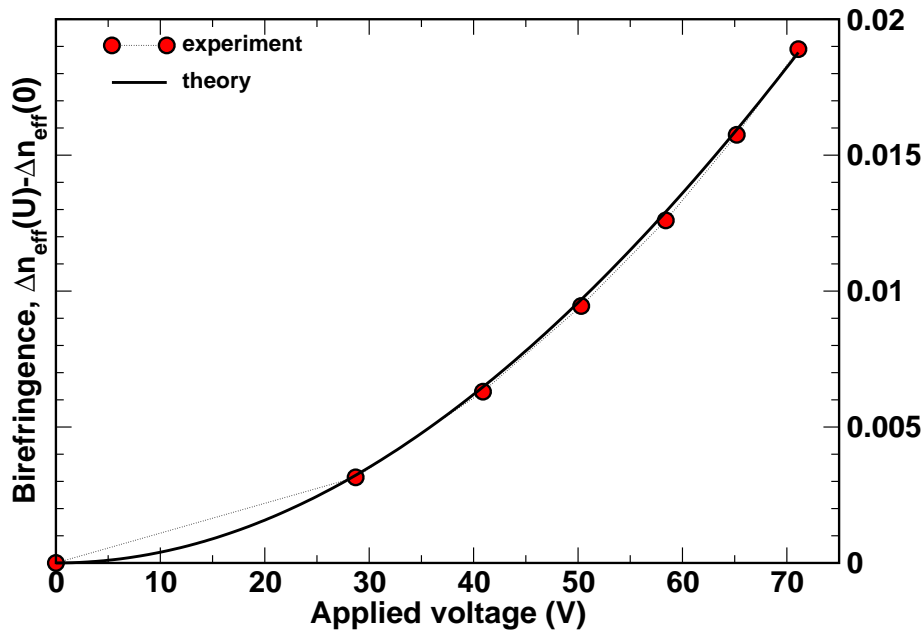}}
\caption{%
Electrically controlled birefringence,
$\Delta n_{\eff}(U)-\Delta n_{\eff}(0)$,
as a function of applied voltage.
The theoretical curve (solid line) is computed from
Eq.~\eqref{eq:epsl-pm}
at $\alpha_{E}=\gamma_{V} U$ and
$\gamma_V\approx 0.0124$~V$^{-1}$.
}
\label{fig:dn_50}
\end{figure*}

Figure~\ref{fig:dn_50} shows the results for 
the voltage dependence of the electrically 
controlled birefringence, 
$\Delta n_{\eff}(U)-\Delta n_{\eff}(0)$,
measured for the mixture FLC-576A.
As can be seen from the figure, there is 
a remarkable accord
between the experimental points 
the curve evaluated from the principal values
of the effective dielectric tensor~\eqref{eq:epsl-eff-diag}.

 In addition to the birefringence,
the transmittance~\eqref{eq:T-xy} depends on the azimuthal angle,
$\phi_{\dd}$, given in Eq.~\eqref{eq:phi-d}.
This angle describes the electric field induced rotation
of the optical axes of the dielectric tensor~\eqref{eq:epsl-eff-diag}
and, under the action of the electric field, $E$, 
its magnitude varies in the range between zero
and $\pi/4$.

A quick inspection of the formula~\eqref{eq:T-xy} shows
that the transmittance will always vanish in 
the zero-field limit with $E=0$.
Such behavior, however, comes as no surprise if we
recall that the electric field of the incident wave
is assumed to be parallel to the helix axis, $\vc{E}_{\inc}\parallel\uvc{x}$,  
which is the zero-voltage optical axis.

Interestingly, the transmittance versus electric field curve
will flatten in the vicinity of the origin
when the cell thickness is chosen in such way 
that, in addition to $\sin\phi_{\dd}$, 
the birefringence dependent factor
equals zero at $E=0$.
The latter is the case for the curves 
shown in Fig.~\ref{fig:data_50}.

\section{Discussion and conclusions}
\label{sec:discussion}

In this paper we have formulated 
the theoretical approach to the optical properties of 
polarization gratings  that can be regarded as
an extension of the method developed in 
Refs~\cite{Kis:jpcm:2007,Kiselev:pra:2008}
for stratified anisotropic media. 
Mathematically, in this approach,
the key formulas~\eqref{eq:Tn-gen} and~\eqref{eq:Rn-gen} 
give 
the transmission and reflection matrices 
of diffraction orders, $\mvc{T}_n$  and $\mvc{R}_n$, 
expressed in terms of the linking matrix~\eqref{eq:W_nk}
which is related to
the evolution operator~\eqref{eq:evol_problem_grt} 
of the system of matrix equations~\eqref{eq:sys-Fn-grt} 
for the Floqu\'et harmonics~\eqref{eq:Fn-grt}
derived
from the Maxwell's equations for the lateral components
of the electric and magnetic fields~\eqref{eq:op-system}.

This method goes beyond the well known limitations of 
the Jones matrix formalism at the expense of simplicity.
To some extent
it can be regarded as a version of the coupled mode analysis
and our
derivation of equations for the Floqu\'et harmonics 
bear some similarity to 
the differential theory of light diffraction~\cite{Neviere:bk:2003}
which is based on the integration of a differential set of equations
derived from Maxwell's equations projected onto some functional bases.

In contrast to a theoretical method 
recently suggested in~\cite{Kreymer:optex:2010}
for a reverse  twisted nematic LC grating,
our approach is formulated without recourse to
the vector theory of scattering for a far field.
Note that the optical properties of  PGs were also analyzed numerically 
using the finite-difference time-domain method
in Ref.~\cite{Escuti:pra:2007}.
Although such analysis is general and useful, it can be very computationally intense
so as to produce the results of sufficiently high accuracy.  

We have used the method described in Sec.~\ref{sec:theory}
as a tool of theoretical investigation into 
the electro-optic properties of 
deformed helix ferroelectric liquid crystal gratings
with subwavelength pitch. 
It was previously demonstrated that a short helix pitch FLC 
(P = 0.4 -0.8~\mum) 
provides an effective phase shift change of a transmitted
polarized light beam as a function of the applied electric field
intensity~\cite{Beresnev:lc:1989}. 
The new mixtures 
with the helix pitch being in
UV region~\cite{Pozhidaev:mclc:2009}
allow to get rid of complications related to the light scattering effects.

So, in such mixtures, we deal with a pure electrically controlled phase
shift plate based on the electro-optical mode called as deformed helix
ferroelectric (DHF). 
DHF is a convenient operation mode able to
ensure both low voltage and fast switching liquid crystalline light
shutters (the response time is less than 200 $\mu$s at driving
electric field around 1 V/\mum). 
For these reasons, it is important
to understand behavior of the electrically controlled
birefringence in such DHFLC cells.

We have shown that, in the short-pitch approximation,
DHFLC cells are equivalent to uniformly anisotropic
biaxial layers with the optical axis normal to the substrate
plane. The latter implies that normally incident light will feel
only uniaxial in-plane anisotropy in agreement with
the recent results on the ellipticity of light 
transmitted through a DHFLC cell~\cite{Pozhidaev:lc:2:2010}.

We have computed the averaged dielectric tensor
as a function of the applied electric field
and have used the results to evaluate 
the light transmittance measured in our experiments.
A comparison between theoretical and experimental results
presented in Figs.~\ref{fig:data_130}--\ref{fig:dn_50}
shows that the predictions of the theory are in good agreement
with the experimental data.
Note that our calculations of the averaged dielectric tensor
rely on the approximate expression for the azimuthal
angle~\eqref{eq:Phi} which is 
applicable for sufficiently low voltages.
A more accurate description of the DHFLC orientational structure
is required  when the voltage increases approaching the unwinding
transition~\cite{Suwa:jjap:2003}. 

\begin{acknowledgments}
This work is supported by HKUST grant CERG 612208, CERG RPC07/08.EG01
and CERG 612409.
A.D.K acknowledges partial financial support under STCU Grant No.~4687.
\end{acknowledgments}

\appendix

\section{Derivation of equations for lateral components}
\label{sec:math}

In this section we discuss how to exclude
the $z$-components of the electromagnetic field,
$E_z$ and $H_z$,
that enter the representation~\eqref{eq:decomp-E},
from the Maxwell equations~\eqref{eq:maxwell}.
Our task is  to derive the closed system of equations
for the lateral (tangential) components,
$\vc{E}_P$ and $\vc{H}_P$.

We begin with
substituting Eqs.~\eqref{eq:decomp-E} and~\eqref{eq:decomp-nabla}
and have   Maxwell's equations~\eqref{eq:maxwell}
recast into the following form:
\begin{subequations}
  \label{eq:maxwell-p-1}
\begin{align}
&
\label{eq:mxwll-Ep-1}
-i\pdrs{\tau}\,
   [\uvc{z}\times\vc{E}_{P}]=\mu  \vc{H}
-
\bs{\nabla}_{p}\times\vc{E},
\\
&
\label{eq:mxwll-Hp-1}
-i\pdrs{\tau}\,
   \vc{H}_{P}=\vc{D}
+
\bs{\nabla}_{p}\times\vc{H},
\end{align}
\end{subequations}
where the explicit expressions for the last terms on the right hand side of the
system~\eqref{eq:maxwell-p-1}
are as follows
\begin{subequations}
  \label{eq:aux-np}
\begin{align}
&
  \label{eq:aux-np-E}
\bs{\nabla}_{p}\times\vc{E}
=
-\bs{\nabla}_{p}^{\perp} E_z+
\sca{\bs{\nabla}_{p}^{\perp}}{\vc{E}_{P}}\,
\uvc{z},
\\
&
\label{eq:aux-np-H}
\bs{\nabla}_{p}\times\vc{H}
=
-\bs{\nabla}_{p}^{\perp} H_z+
\sca{\bs{\nabla}_{p}}{\vc{H}_{P}}\,
\uvc{z}.
\end{align}
\end{subequations}

We can now substitute 
the electric displacement field written as a sum of
the normal and in-plane components
\begin{align}
\label{eq:decomp-D}
  \vc{D}=D_z \uvc{z} +\vc{D}_{P},  
\end{align}
into Eq.~\eqref{eq:mxwll-Hp-1}
and derive the following expression for
its $z$-component
\begin{align}
  &
\label{eq:D_z}
D_z=
\epsilon_{zz} E_z+
\sca{\bs{\epsilon}_{z}}{\vc{E}_{P}}
=
-\sca{\bs{\nabla}_{p}}{\vc{H}_{P}},
\end{align}
where $\bs{\epsilon}_{z}=(\epsilon_{zx},\epsilon_{zy})$.

From Eqs.~\eqref{eq:maxwell-p-1}
and~\eqref{eq:D_z},
it is not difficult to deduce the relations
\begin{align}
&
   \label{eq:H_z}
H_z=
\mu^{-1}
\sca{\bs{\nabla}_{p}^{\perp}}{\vc{E}_{P}}
\\
&
\label{eq:E_z}
 E_z
=
-\epsilon_{zz}^{-1}
\bigl[
\sca{\bs{\epsilon}_{z}}{\vc{E}_{P}}
+\sca{\bs{\nabla}_{p}}{\vc{H}_{P}}
\bigr]
\end{align}
linking the normal (along the $z$ axis) and  the lateral
(perpendicular to the $z$ axis) components. 

By using the relation~\eqref{eq:E_z},
we obtain
the tangential component of the field~\eqref{eq:decomp-D}
\begin{align}
&
\label{eq:D_p}
\vc{D}_{P}=
\bs{\epsilon}_{z}^{\,\prime} E_z
+
\bs{\varepsilon}_z\cdot \vc{E}_{P}
=
\bs{\varepsilon}_{P}\cdot \vc{E}_{P}
-
\bs{\epsilon}_{z}^{\,\prime}\,
\epsilon_{zz}^{-1}
\sca{\bs{\nabla}_{p}}{\vc{H}_{P}}
\end{align}
where
$
\bs{\varepsilon}_z=
\begin{pmatrix}
  \epsilon_{xx}& \epsilon_{xy}\\
\epsilon_{yx} & \epsilon_{yy} 
\end{pmatrix}$;
$\bs{\epsilon}_{z}^{\,\prime}=(\epsilon_{xz},\epsilon_{yz})$
and the effective dielectric tensor, $\bs{\varepsilon}_{P}$, 
for the lateral components is given by
\begin{align}
  \label{eq:diel-p}
 \bs{\varepsilon}_{P}
=
 \bs{\varepsilon}_{z}-
\epsilon_{zz}^{-1}\,
\bs{\epsilon}_{z}^{\,\prime}\otimes
\bs{\epsilon}_{z}.
\end{align}

Maxwell's equations~\eqref{eq:maxwell-p-1} 
can now be combined with the relations~\eqref{eq:aux-np}
to yield the system
\begin{subequations}
  \label{eq:maxwell-p-2a}
\begin{align}
&
\label{eq:mxwll-Ep-2a}
-i\pdrs{\tau}\,
   \vc{E}_{P}= \mu  \vc{H}_{P}
+ \bs{\nabla}_{p} E_z,
\\
&
\label{eq:mxwll-Hp-2a}
-i\pdrs{\tau}\,
   \vc{H}_{P}=\vc{D}_P
-
\bs{\nabla}_{p}^{\perp} H_z,
\end{align}
\end{subequations}
where $H_z$, $E_z$
and $\vc{D}_P$ are given in
Eq.~\eqref{eq:H_z},
Eq.~\eqref{eq:E_z} and
Eq.~\eqref{eq:D_p}, respectively.

So, this system immediately gives the final result
\begin{subequations}
  \label{eq:maxwell-p-2}
\begin{align}
&
  \label{eq:mxwll-Ep-2}
-i\pdrs{\tau}\,
   \vc{E}_{P}=
-
\bs{\nabla}_{p}
[\epsilon_{zz}^{-1}\sca{\bs{\epsilon}_{z}}{\vc{E}_{P}}]
+
\mu  \vc{H}_{P}
-
\bs{\nabla}_{p}
[\epsilon_{zz}^{-1}\sca{\bs{\nabla}_{p}}{\vc{H}_{P}}],
\\
&
\label{eq:mxwll-Hp-2}
-i\pdrs{\tau}\,
   \vc{H}_{P}=
\bs{\varepsilon}_{P}\cdot \vc{E}_{P}
-
\bs{\nabla}_{p}^{\perp}
[\sca{\bs{\nabla}_{p}^{\perp}}{\vc{E}_{P}}/\mu]
-
\bs{\epsilon}_{z}^{\,\prime}\,
\epsilon_{zz}^{-1}
\sca{\bs{\nabla}_{p}}{\vc{H}_{P}}
\end{align}
\end{subequations}
that can be easily rewritten in the matrix form~\eqref{eq:op-system}
used in Sec.~\ref{sec:theory}.

\section{Stratified anisotropic media}
\label{sec:strat-anis-media}

In this appendix we 
the limiting case of stratified medium
with vanishing modulation of the dielectric
 tensor~\eqref{eq:diel-Four-grt}, so that
$\bs{\varepsilon}=\bs{\varepsilon}_{0}$.
Then the only mode of interest
is the zero-order harmonics
$\vc{F}=\vc{F}_0$
governed by the matrix equation
\begin{align}
  \label{eq:sys-F-strat}
   -i\pdrs{\tau}\vc{F}(\tau)=
\mvc{M}(\tau)\cdot \vc{F}(\tau),
\end{align}
where $\mvc{M}=\mvc{M}_{00}$.

The evolution operator of Eq.~\eqref{eq:sys-F-strat}
is defined as follows
\begin{align}
  \label{eq:evol_problem_eq}
  -i\pdrs{\tau}\mvc{U}(\tau,\tau_0)=
\mvc{M}(\tau)\cdot \mvc{U}(\tau,\tau_0),
\quad
\mvc{U}(\tau_0,\tau_0)=\mvc{I}_4.
\end{align}
We now
briefly touch on some of its basic properties. 

\subsection{Operator of evolution}
\label{subsec:op-evol}

We begin with the relation 
\begin{align}
  \label{eq:compos_law}
  \mvc{U}(\tau,\tau_0)=\mvc{U}(\tau,\tau_1)\cdot \mvc{U}(\tau_1,\tau_0)
\end{align}
known as the \textit{composition law}.
This result derives from the fact that the operator
$\mvc{U}(\tau,\tau_0)\cdot \mvc{U}^{-1}(\tau_1,\tau_0)$
is the solution of the initial value problem~\eqref{eq:evol_problem_eq}
with $\tau_0$ replaced by $\tau_1$.

From the composition law~\eqref{eq:compos_law}
it immeadiately follow that 
the inverse of the evolution operator is given by
\begin{align}
  \label{eq:inv_evol_oper_1}
\mvc{U}^{-1}(\tau,\tau_0)=    
\mvc{U}(\tau_0,\tau)
\end{align}
and can be found by solving the initial value problem
\begin{align}
  \label{eq:eq_inv_evoper}
     i\pdrs{\tau}\mvc{U}^{-1}(\tau,\tau_0)=\mvc{U}^{-1}(\tau,\tau_0)\cdot
     \mvc{M}(\tau),
\quad
\mvc{U}^{-1}(\tau_0,\tau_0)=\mvc{I}_4.
\end{align}

For non-absorbing media with symmetric dielectric tensor, 
the matrix $\mvc{M}$ is real-valued, $\cnj{\mvc{M}}=\mvc{M}$,
and meets the following symmetry identities~\cite{Kiselev:pra:2008}:
\begin{align}
  \label{eq:symm_M}
  \cnj{\mvc{M}}=\mvc{M},\quad
\tcnj{(\mvc{G}\cdot\mvc{M})}=
\mvc{G}\cdot\mvc{M},
\quad
\mvc{G}=
\begin{pmatrix}
  \mvc{0}&\mvc{I}_2\\
\mvc{I}_2&\mvc{0}
\end{pmatrix},
\end{align}
where an asterisk and the superscript $T$ indicate 
complex conjugation and
matrix transposition, respectively. 
In this case, the evolution operator and its inverse
are related as follows: 
\begin{align}
  \label{eq:unit_U}
  \mvc{U}^{-1}(\tau,\tau_0)
=
\mvc{G}\cdot
\hcnj{\mvc{U}}(\tau,\tau_0)
\cdot
\mvc{G},
\end{align}
where a dagger will denote Hermitian conjugation.
By using the relations~\eqref{eq:symm_M},
it is not difficult to verify that
the operator on the right hand side of Eq.~\eqref{eq:unit_U}
is the solution of the Cauchy problem~\eqref{eq:eq_inv_evoper}. 

We are now in position to deduce
the unitarity relation 
\begin{align}
  \label{eq:unit_W}
  \mvc{W}^{-1}
=
\mvc{G}_3\cdot
\hcnj{\mvc{W}}
\cdot
\mvc{G}_3
=
\begin{pmatrix}
\hcnj{\mvc{W}}_{11} & -\hcnj{\mvc{W}}_{21}\\
-\hcnj{\mvc{W}}_{12} & \hcnj{\mvc{W}}_{22}
\end{pmatrix}
,
\quad
\mvc{G}_3=\diag(\mvc{I}_2,-\mvc{I}_2)
\end{align}
for the linking matrix
\begin{align}
\label{eq:W-op}
  \mvc{W} = \mvc{V}_{\med}^{-1}\cdot\mvc{U}^{-1}(h,0)\cdot \mvc{V}_{\med}
\end{align}
by using Eq.~\eqref{eq:unit_U}
in combination with
the algebraic identity
\begin{align}
  \label{eq:idnt_Vm}
\tcnj{\mvc{V}}_{\med}
\cdot
\mvc{G}\cdot\mvc{V}_{\med}=
N_{\med}\mvc{G}_3,
\end{align}
where $N_{\med} = 2 q_{\med}/\mu_{\med}$,
for the eigenvector matrix
\begin{align}
\label{eq:Vm-block}
\mvc{V}_{\med}\equiv\mvc{V}_{\med}(q_\med)=
\begin{pmatrix}
\mvc{E}_{\med} & -\bs{\sigma}_3 \mvc{E}_{\med}\\
\mvc{H}_{\med} & \bs{\sigma}_3 \mvc{H}_{\med}\\
\end{pmatrix}  
\end{align}
with $\mvc{E}_{\med}=\diag(q_{\med}/n_{\med},1)$
and $\mu_{\med}\,\mvc{H}_{\med}=\diag(n_{\med},q_{\med})$. 

\begin{figure*}[!tbh]
\centering
  \resizebox{90mm}{!}{\includegraphics*{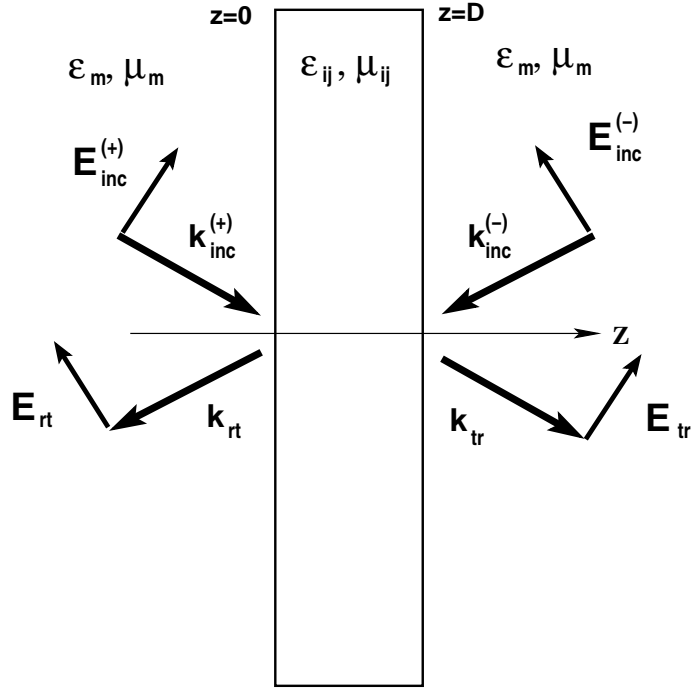}}
\caption{%
Four-wave geometry with two incident waves,
$\vc{E}_{\inc}^{(+)}$ and $\vc{E}_{\inc}^{(-)}$,
impinging onto the entrance ($z=0$) and exit ($z=D$) 
faces, respectively.
}
\label{fig:geom_4rays}
\end{figure*}

\subsection{Linking Matrix}
\label{subsec:link-matr}

For closer examination of properties of the linking matrix~\eqref{eq:W-op}, 
we consider the four-wave geometry 
shown in Fig.~\ref{fig:geom_4rays},
where $\vc{k}_{\inc}^{(+)}=\vc{k}_{\ind{tr}}=\vc{k}_{\inc}=\vc{k}_{\transm}$
and $\vc{k}_{\inc}^{(-)}=\vc{k}_{\ind{rt}}=\vc{k}_{\refl}$.
In this geometry, there are two plane waves, 
$\vc{E}_{\inc}^{(+)}$ and $\vc{E}_{\inc}^{(-)}$,
incident on both bounding surfaces of the anisotropic layer. 
For the transmitted and reflected waves,
 $\vc{E}_{\ind{tr}}$ and $\vc{E}_{\ind{rt}}$,
excited by the incident light,
the input-output relation~\eqref{eq:transm-rel} 
takes the following generalized form:
\begin{align}
  \label{eq:TR_gen}
    \begin{pmatrix}
\vc{E}_{\ind{tr}}\\
\vc{E}_{\ind{rt}}
\end{pmatrix}
=
\begin{pmatrix}
\mvc{T}_{+} & \mvc{R}_{-}\\
\mvc{R}_{+} & \mvc{T}_{-}
\end{pmatrix}
\cdot
\begin{pmatrix}
\vc{E}_{\inc}^{(+)}\\
\vc{E}_{\inc}^{(-)}
\end{pmatrix},
\end{align}
where $\mvc{T}_{+}$ ($\mvc{R}_{+}$) is the transmission (reflection)
matrix defined in Eq.~\eqref{eq:transm-rel},
whereas the mirror symmetric case where the incident wave is impinging onto 
the exit face of the sample is described by
the transmission (reflection) matrix $\mvc{T}_{-}$ ($\mvc{R}_{-}$).

The matrix~\eqref{eq:W-op} is introduced through the relation~\cite{Kis:jpcm:2007,Kiselev:pra:2008}
\begin{align}
  \label{eq:W_def}
\mvc{W}
    \begin{pmatrix}
\vc{E}_{\ind{tr}}\\
\vc{E}_{\inc}^{(-)}
\end{pmatrix}
=
\begin{pmatrix}
\vc{E}_{\inc}^{(+)}\\
\vc{E}_{\ind{rt}}
\end{pmatrix}
\end{align}
linking the electric field vector amplitudes of the waves
in the half spaces $z<0$ and $z>D$
bounded by the faces of the layer.
From Eqs.~\eqref{eq:TR_gen} and~\eqref{eq:W_def}, its block structure
expressed in terms of the transmission and reflection matrices
is as follows  
\begin{align}
  \label{eq:W_TR}
  \mvc{W}=
\begin{pmatrix}
\mvc{T}_{+}^{-1} & -\mvc{T}_{+}^{-1}\cdot\mvc{R}_{-}\\
\mvc{R}_{+}\cdot\mvc{T}_{+}^{-1} & \mvc{T}_{-}-\mvc{R}_{+}\cdot\mvc{T}_{+}^{-1}\cdot\mvc{R}_{-}
\end{pmatrix}.
\end{align}
Similarly, for inverse of the linking matrix,
we have
\begin{align}
  \label{eq:inv_W_TR}
  \mvc{W}^{-1}=
\begin{pmatrix}
\mvc{W}_{11}^{(-1)}& \mvc{W}_{12}^{(-1)} \\
\mvc{W}_{21}^{(-1)} & \mvc{W}_{22}^{(-1)}
\end{pmatrix}
=
\begin{pmatrix}
\mvc{T}_{+}-\mvc{R}_{-}\cdot\mvc{T}_{-}^{-1}\cdot\mvc{R}_{+} & \mvc{R}_{-}\cdot\mvc{T}_{-}^{-1}\\ 
 -\mvc{T}_{-}^{-1}\cdot\mvc{R}_{+}& \mvc{T}_{-}^{-1} 
\end{pmatrix}.
\end{align}

We can now use the unitarity relation~\eqref{eq:unit_W} for non-absorbing layers
to derive 
the energy conservation law
\begin{align}
  \label{eq:energy_consv}
  \hcnj{\mvc{T}}_{\pm}\cdot\mvc{T}_{\pm}+ 
\hcnj{\mvc{R}}_{\pm}\cdot\mvc{R}_{\pm}=\mvc{I}_2.
\end{align}
along with  the relations for the block matrices 
\begin{subequations}
\label{eq:W_ij_uni}
\begin{align}
&
  \label{eq:W_ii_uni}
  \mvc{W}_{11}=\mvc{T}_{+}^{-1},\quad
  \mvc{W}_{22}=\hcnj{[\mvc{T}_{-}^{-1}]},
\\
&
 \label{eq:W_12_uni}
  \mvc{W}_{12}=-\mvc{T}_{+}^{-1}\cdot\mvc{R}_{-}=
\hcnj{[\mvc{T}_{-}^{-1}\cdot\mvc{R}_{+}]},
\\
&
 \label{eq:W_21_uni}
  \mvc{W}_{21}=\mvc{R}_{+}\cdot\mvc{T}_{+}^{-1}=
-\hcnj{[\mvc{R}_{-}\cdot\mvc{T}_{-}^{-1}]}.
\end{align}
\end{subequations}

In the translation invariant case of uniform anisotropy, 
the matrix $\mvc{M}$
is independent of $\tau$ and
the operator of evolution is given by
\begin{align}
  \label{eq:U-hom}
  \mvc{U}(\tau,\tau_0)=
\mvc{U}(\tau-\tau_0)=
     \exp\{i \mvc{M}\, (\tau-\tau_0)\}.
\end{align} 
Then, the unitarity condition~\cite{Kiselev:pra:2008}
\begin{align}
  \label{eq:unit_UW-hom}
  \mvc{U}^{-1}=\cnj{\mvc{U}},
\quad
\mvc{W}^{-1}=\cnj{\mvc{W}}
\end{align}
can be combined with Eq.~\eqref{eq:unit_W} 
to yield the additional symmetry relations
for $\mvc{W}_{ij}$
\begin{align}
  \label{eq:W_ij_hom}
  \tcnj{\mvc{W}}_{ii}=\mvc{W}_{ii},
\quad
  \tcnj{\mvc{W}}_{12}=-\mvc{W}_{21}
\end{align}
that give the following algebraic identities for the transmission and
reflection matrices:
\begin{align}
&
  \label{eq:TR-sym-hom}
   \tcnj{\mvc{T}}_{\pm}=\mvc{T}_{\pm},
\quad
   \tcnj{\mvc{R}}_{+}=\mvc{R}_{-},
\\
&
  \label{eq:T-sym-hom}
  \cnj{\mvc{T}}_{\pm}=-\cnj{\mvc{R}}_{\mp}
\cdot\mvc{T}_{\mp}\cdot\mvc{R}_{\mp}^{-1}.
\end{align}

\subsection{Uniformly anisotropic planar structure}
\label{subsec:planar}

In this section we present the results for
the planar oriented nematic cell. 
For the planar orientational structure,
the director is given by
\begin{align}
  \label{eq:director_pln}
 \uvc{d}=(d_x,d_y,d_z)=(\cos\phi_{\dd},\sin\phi_{\dd},0), 
\end{align}
so that the layer is uniformly anisotropic.
 
Assuming that the plane of incidence
is defined as the $x$-$z$ plane ($\phi_{\inc}=0$),
we write down 
the elements of the matrix $\mvc{M}$
as follows 
\begin{align}
  \label{eq:matrix-M12}
  &
\mvc{M}_{12}=\mu\mvc{I}_2-\frac{q_x^2}{2\epsilon_{\perp}}
(\mvc{I}_2+\bs{\sigma}_3),
\quad
\mvc{M}_{ii}=\mvc{0},
\\
&
  \label{eq:matrix-M21}
 \mvc{M}_{21}=
-\frac{q_x^2}{2\mu}
(\mvc{I}_2-\bs{\sigma}_3)
+\epsilon_c
\Bigl\{
\mvc{I}_2+ 
\tilde{u}_a \bigl[
\cos(2\phi_{\dd})\,\bs{\sigma}_3+
\sin(2\phi_{\dd})\,\bs{\sigma}_1
\bigr]
\Bigr\},
\\
 \label{eq:epsl_c}
&
  \epsilon_c=(\epsilon_{\parallel}+\epsilon_{\perp})/2,
\quad
\tilde{u}_a=\frac{\epsilon_{\parallel}-\epsilon_{\perp}}{\epsilon_{\parallel}+\epsilon_{\perp}},
\quad
q_{x}\equiv q_x^{(p)},
\end{align}
where  the azimuthal angle of the in-plane optical axis $\phi_{\dd}$ 
also gives the angle between
the director and the plane of incidence.

For uniform anisotropy, this matrix is constant.
Therefore, the operator of evolution
can be expressed in terms of 
the eigenvalue and eigenvector matrices, 
$\mvc{\Lambda}$ and $\mvc{V}$, as follows
\begin{align}
  \label{eq:U-pln}
  \mvc{U}(h)=
     \exp\{i \mvc{M}\, h\}=\mvc{V}\cdot
      \exp\{i \mvc{\Lambda}\, h\}
\cdot
\mvc{V}^{-1}.
\end{align}
It is not difficult to 
solve the eigenvalue problem for the matrix $\mvc{M}$
and find the expressions for the eigenvalues
that enter the eigenvalue matrix
\begin{align}
&
  \label{eq:Q_N-pln}
\mvc{\Lambda}=\diag(\mvc{Q},-\mvc{Q}),
\quad
\mvc{Q}=\diag(q_e,q_o),
\\
&
  \label{eq:qz_eo-pln}
  q_{e}=\sqrt{n_e^2-q_x^2(1+u_a d_x^2)},
\quad
  q_{o}=\sqrt{n_o^2-q_x^2},
\end{align}
where $u_a=\Delta\epsilon/\epsilon_{\perp}$. 
Similarly, after computing the eigenvectors,
we obtain the eigenvector matrix in the following form:
\begin{align}
&
\label{eq:V-pln}
\mvc{V}=
\begin{pmatrix}
            \mvc{E} & \mvc{E}\cdot\bs{\sigma}_3 \\
             \mvc{H} & -\mvc{H}\cdot\bs{\sigma}_3
\end{pmatrix},
\\
&
\label{eq:E-pln}
\mvc{E}=\mu\,
\begin{pmatrix}
            d_x[1-q_x^2/n_o^2] & d_yq_o \\
             d_y & -d_xq_o
\end{pmatrix},
\\
&
\label{eq:H-pln}
\mvc{H}=
\begin{pmatrix}
              d_x q_e & d_y n_o^2 \\
             d_y q_e & -d_x [n_o^2-q_x^2]
\end{pmatrix}.
\end{align}
Upon substituting Eqs.~\eqref{eq:U-pln}--~\eqref{eq:H-pln}
into Eq.~\eqref{eq:W-op}, some rather 
straightforward algebraic manipulations give the linking matrix
\begin{align}
&
\label{eq:W-pln}
  \mvc{W}=N_\med^{-1}\diag(\mvc{I}_2,\bs{\sigma}_3)\cdot
\tilde{\mvc{W}}\cdot\diag(\mvc{I}_2,\bs{\sigma}_3),
\\
&
 \label{eq:tW-pln}
\tilde{\mvc{W}}=
\begin{pmatrix}
  \mvc{A}_{+} & \mvc{A}_{-}\\
\mvc{A}_{-} & \mvc{A}_{+}
\end{pmatrix}
\cdot
\mvc{W}_\dd
\cdot
\begin{pmatrix}
 \mvc{A}_{+}^{T} &  -\mvc{A}_{-}^{T} \\
 -\mvc{A}_{-}^{T} &  \mvc{A}_{+}^{T}
\end{pmatrix}
\\
&
\label{eq:Wd-pln}
\mvc{W}_\dd=
\begin{pmatrix}
  \mvc{W}_{-} & \vc{0}\\
\vc{0} & \mvc{W}_{+}
\end{pmatrix},
\quad
  \mvc{W}_{\pm}=\exp[\pm i \mvc{Q} h]\cdot\mvc{N}^{-1},
\\
&
  \label{eq:A_pm-pln}
  \mvc{A}_{\pm}=\mvc{E}_{\med}\cdot\mvc{H}\pm\mvc{H}_{\med}\cdot\mvc{E},
\quad
\mvc{N}=\diag(N_e,N_o),
\\
&
\label{eq:N_oe-pln}
N_{e}=\dfrac{2q_e\mu}{n_o^2}(n_o^2-q_x^2d_x^2),
\quad
N_{o}=2q_o\mu (n_o^2-q_x^2d_x^2),
\end{align}
where $N_{\med}=2 q_{\med}/\mu_{\med}$.
From Eq.~\eqref{eq:tW-pln}, 
the block $2\times 2$ matrices of $\tilde{\mvc{W}}$
are given by
\begin{subequations}
\label{eq:tW_ij-pln}
\begin{align}
&
\label{eq:tW_11-pln}
\tilde{\mvc{W}}_{11}=
N_{\med}\mvc{W}_{11}=
\mvc{A}_{+}\cdot\mvc{W}_{-}\cdot\mvc{A}_{+}^{T}-\mvc{A}_{-}\cdot\mvc{W}_{+}\cdot\mvc{A}_{-}^{T},
\\
&
\label{eq:tW_22-pln}
\tilde{\mvc{W}}_{22}=
N_{\med}\bs{\sigma}_3\cdot\mvc{W}_{22}\cdot\bs{\sigma}_3=
\mvc{A}_{+}\cdot\mvc{W}_{+}\cdot\mvc{A}_{+}^{T}-\mvc{A}_{-}\cdot\mvc{W}_{-}\cdot\mvc{A}_{-}^{T},
\\
&
\label{eq:tW_21-pln}
\tilde{\mvc{W}}_{21}=
N_{\med}\bs{\sigma}_3\cdot\mvc{W}_{21}=
-\tcnj{\tilde{\mvc{W}}}_{12}=
-N_{\med}\tcnj{[\mvc{W}_{12}\cdot\bs{\sigma}_3]}=
\notag
\\
&
=\mvc{A}_{-}\cdot\mvc{W}_{-}\cdot\mvc{A}_{+}^{T}-\mvc{A}_{+}\cdot\mvc{W}_{+}\cdot\tcnj{\mvc{A}}_{-}.  
\end{align}
\end{subequations}

Finally, we can combine 
Eq.~\eqref{eq:W_TR}
with Eq.~\eqref{eq:W-pln} to derive
the expressions for the transmission and reflection matrices
\begin{align}
&
\label{eq:TR_pln}
\mvc{T}_{+}=N_{\med}
\tilde{\mvc{W}}_{11}^{-1},
\quad
\mvc{R}_{+}= \bs{\sigma}_3\cdot\tilde{\mvc{W}}_{21}\cdot\tilde{\mvc{W}}_{11}^{-1}.
\end{align}

As it can be seen from the formulas~\eqref{eq:tW_ij-pln},
the symmetry relations~\eqref{eq:W_ij_hom} are satisfied.
Interestingly, when the eigenvalue matrix~\eqref{eq:Q_N-pln}
is real so that $\hcnj{\mvc{W}}_{+}=\mvc{W}_{-}$
and $\hcnj{\mvc{A}}_{\pm}=\tcnj{\mvc{A}}_{\pm}$, 
close inspection of the expressions~\eqref{eq:tW_ij-pln}
shows that, in addition to the symmetry relations
for uniform anisotropy~\eqref{eq:W_ij_hom},
the case of uniform planar structure
is characterized by the following algebraic identities:
\begin{align}
  \label{eq:Wij_hsym_pln}
  \hcnj{\mvc{W}}_{22}=
\bs{\sigma}_3\cdot\mvc{W}_{11}\cdot\bs{\sigma}_3,
\quad
  \hcnj{\mvc{W}}_{21}=
-\bs{\sigma}_3\cdot\mvc{W}_{21}\cdot\bs{\sigma}_3.
\end{align}
These identities and the unitarity conditions~\eqref{eq:W_ij_uni}
can now be used to deduce the relations 
\begin{align}
  \label{eq:TR_hsym_pln}
  \mvc{T}_{+}=
\bs{\sigma}_3\cdot\mvc{T}_{-}\cdot\bs{\sigma}_3,
\quad
  \mvc{R}_{+}=
\bs{\sigma}_3\cdot\mvc{R}_{-}\cdot\bs{\sigma}_3
\end{align}
linking the transmission (reflection) matrix,
$\mvc{T}_{+}\equiv \mvc{T}$ ($\mvc{R}_{+}\equiv \mvc{R}$),
and its mirror symmetric counterpart
 $\mvc{T}_{-}$ ($\mvc{R}_{-}$).

Before closing this section we briefly comment on 
the important special case of normal incidence
that occurs at $q_x=0$.
In this case, the matrices $\mvc{A}_{\pm}$ defined in Eq.~\eqref{eq:A_pm-pln}
can be written in the factorized form
\begin{align}
  \label{eq:Apm_norm_pln}
  \mvc{A}_{\pm}(\phi_{\dd})=\mvc{Rt}(\phi_{\dd})\cdot
\mvc{A}_{\pm}(0)=
\begin{pmatrix}
  d_x & -d_y\\
d_y& d_x
\end{pmatrix}
\cdot
\begin{pmatrix}
  \dfrac{\mu_{\med}\,n_e\pm \mu\, n_{\med}}{\mu_{\med}}&0\\
0& -n_o\dfrac{\mu_{\med}\,n_o\pm \mu\, n_{\med}}{\mu_{\med}}
\end{pmatrix},
\end{align}
where $\mvc{Rt}(\phi)=\begin{pmatrix}
  \cos\phi &-\sin\phi\\
\sin\phi & \cos\phi
\end{pmatrix}$ 
is the matrix describing rotation
about the $z$ axis by the angle $\phi$.
Substituting Eq.~\eqref{eq:Apm_norm_pln}
into  Eq.~\eqref{eq:tW_ij-pln}
gives the block matrices  
\begin{align}
  \label{eq:tWij_norm_pln}
  \tilde{\mvc{W}}_{ij}(\phi_{\dd})=\mvc{Rt}(\phi_{\dd})\cdot
\tilde{\mvc{W}}_{ij}(0)\cdot\mvc{Rt}(-\phi_{\dd})
\end{align}
expressed as a function of the director azimuthal
angle $\phi_{\dd}$.

The result for the transmission and reflection matrices
\begin{align}
  \label{eq:TR_norm_pln}
  \mvc{T}_{\pm}(\phi_{\dd})=\mvc{Rt}(\pm\phi_{\dd})\cdot
\mvc{T}(0)\cdot\mvc{Rt}(\mp\phi_{\dd}),
\quad
  \mvc{R}_{\pm}(\phi_{\dd})=\mvc{Rt}(\mp\phi_{\dd})\cdot
\mvc{R}(0)\cdot\mvc{Rt}(\mp\phi_{\dd}),
\end{align}
where the diagonal matrices $\mvc{T}(0)=\diag(t_e,t_o)$ and $\mvc{R}(0)=\diag(r_e,-r_o)$
\begin{align}
&
  \label{eq:tr_norm_plm}
  t_{\alpha}=\frac{1-\rho_{\alpha}^2}{%
1-\rho_{\alpha}^2\exp(2in_{\alpha}h)
}\exp(i n_{\alpha} h),
\quad
  r_{\alpha}=\frac{1-\exp(2in_{\alpha}h)}{%
1-\rho_{\alpha}^2\exp(2in_{\alpha}h)
}\rho_{\alpha},
\\
&
\label{eq:rho_norm_pln}
\rho_{\alpha}=\frac{n_{\alpha}/\mu-n_{\med}/\mu_{\med}}{%
n_{\alpha}/\mu+n_{\med}/\mu_{\med}},
\quad
\alpha\in\{e, o\}
\end{align}
describe the  case in which the director~\eqref{eq:director_pln} 
lies in the incidence plane, 
immediately follows from the relations~\eqref{eq:TR_pln}
and~\eqref{eq:TR_hsym_pln}.
Finally, the expressions for the matrices~\eqref{eq:TR_norm_pln}
\begin{align}
&
  \label{eq:T_norm_pln}
  \mvc{T}_{\pm}(\phi_{\dd})=
\frac{t_e+t_0}{2}\,
\mvc{I}_2+\frac{t_e-t_0}{2}\,\
\mvc{Rt}(\pm 2 \phi_{\dd})\cdot
\bs{\sigma}_3,
\\
&
  \label{eq:R_norm_pln}
  \mvc{R}_{\pm}(\phi_{\dd})=
\frac{r_e+r_0}{2}\,
\bs{\sigma}_3 +
\frac{r_e-r_0}{2}\,
\mvc{Rt}(\mp 2 \phi_{\dd})
\end{align}
can be readily obtained by using the identity:
$\bs{\sigma}_3\cdot\mvc{Rt}(\phi)\cdot\bs{\sigma}_3=\mvc{Rt}(-\phi)$.


%

\end{document}